\documentclass[a4paper, twoside]{report}

\usepackage[english]{babel}
\usepackage[utf8x]{inputenc}
\usepackage[T1]{fontenc}

\usepackage[a4paper,top=3cm,bottom=2cm,left=3cm,right=3cm,marginparwidth=2cm]{geometry}

\usepackage{amsmath}
\usepackage{physics}
\usepackage{amsfonts} 
\usepackage{graphicx}
\usepackage{indentfirst}
\usepackage{caption}
\usepackage{subcaption}
\usepackage[colorinlistoftodos]{todonotes}
\usepackage[colorlinks=true, allcolors=blue]{hyperref}

\title{Supergravity and p-brane Ansatz}
\author{Yuelin Shen}

\begin{document}
\begin{titlepage}

\newcommand{\HRule}{\rule{\linewidth}{0.5mm}} 


 

\center 


\textsc{\LARGE Msci Project}\\[1.5cm] 
\textsc{\Large Imperial College London}\\[0.5cm] 
\textsc{\large  Department of Physics, Theoretical Physics Group}\\[0.5cm] 

\makeatletter
\HRule \\[0.4cm]
{ \huge \bfseries \@title}\\[0.4cm] 
\HRule \\[1.5cm]
 

\begin{minipage}{0.4\textwidth}
\begin{flushleft} \large
\emph{Author:}\\
\@author 
\end{flushleft}
\end{minipage}
~
\begin{minipage}{0.4\textwidth}
\begin{flushright} \large
\emph{Supervisor:} \\
Prof. Kellogg Stelle \\[1.2em] 
\emph{Second Marker:} \\
Prof. Daniel Waldram 
\end{flushright}
\end{minipage}\\[2cm]
\makeatother



{\large March 2, 2022}\\[2cm] 
{\large Word Count: 9641}\\[2cm] 
\vfill 

\end{titlepage}

\newgeometry{left=2.5cm,right=2.5cm,top=2cm,bottom=2cm}                          \renewcommand{\abstractname}{\large Acknowledgement}
\begin{abstract}
Special acknowledgement to my supervisor Professor Kellogg Stelle, his PHD student Rahim Leung and the fellows in the project group.
\end{abstract}

\renewcommand{\abstractname}{Abstract}
\begin{abstract}
This project explores the $D=11$ supergravity model and the properties of its p-brane ansatz. The initial field content (graviton, gravitino and the anti-symmetric tensor field) in the action of $D=11$ supergravity is explained in the context of supersymmetry. The action is then decomposed to the bosonic sector, which is compared with the $\sigma$-model in string theory at a low energy limit $\alpha'\xrightarrow{}0$. The dilaton in the $D=10$ string theory can be realised from the dimensional reduction of $D=11$ supergravity, which gives the scalar contribution in the action to form the single-charge action. The field equation of the single-charged action is then derived. An $SO(D-d)\times Poincare_{d}$ ansatz is introduced to simplify the field equation. The solution of the field equation bifurcates into the electric ansatz and the magnetic ansatz. These ansatzes are called p-branes which are string-like objects that exist in their p-dimensional world volume embedded in the ambient spacetime. The BPS bounds are saturated for these p-branes, and upon dimensional reduction, they are similar to extremal Riessner-Nordstrom black holes up to the scalar. The branic motion is then derived and a special case of parallel brane orbit is explored. Similar to the Riessner-Nordstrom black hole, the circular orbit is found to require a specific angular momentum that increases further from the central brane.  The circular orbit always exists for the extremal case, but the black branes that do not saturate the BPS bound may not have a circular orbit below a threshold angular momentum.
\end{abstract}

\tableofcontents
\listoffigures

\chapter{Introduction}
The unification between gravity and quantum mechanics requires a field theory that can localise the fields in the standard model on a curved manifold. The localisation requires more field content underpinned by a larger symmetry group --- supersymmetry,
Supergravity is a theory that incorporates supersymmetry and gravity, which is an important step toward grand unification. \cite{tanii_2014_1}
 
Broadly speaking, supersymmetry is a symmetry between bosons and fermions. It was first introduced to combine the $Poincar\Acute{e}$ symmetry, $ISO(3,1)$, with the internal symmetries in the standard model, such as $SU(2)$.  Supergravity is a supersymmetric theory containing gravity or can be interpreted as a local theory of supersymmetry. 
To have a local theory, one needs to turn the supersymmetric parameter $\epsilon^{\alpha}$ into a local parameter. In the localisation of $U(1)$ symmetry, a gauge field $A_{\mu}$ is required to make the action, $\mathcal{L}=\int \partial_{\mu}\phi\partial^{\mu}\phi$, invariant under the transformation. Similarly, the gauge field $A_{\mu}^{\alpha}$ is required for the local parameter $\epsilon^{\alpha}$. $A_{\mu}^{\alpha}$ can be written as $\psi_{\mu\alpha}$, which is known as the gravitino. The gravitino $\psi_{\mu\alpha}$ is a spinor object with a curved space index $\mu$, thus, its supermultiplet partner should be a bosonic object with a curved space index $\mu$, i.e. $\psi_{\mu\alpha}=Q_{\alpha}\mbox({boson})_{\mu}$. To have gravity as an emergent phenomenon from the supersymmetry, the boson also has to be related to the metric of the curved spacetime. Based on these properties, one can infer that the minimal field content in supergravity should include the gravitino, $\psi_{\mu\alpha}$, and the vielbein, $e_{\mu}^{a}$. \cite{nastase_2015_supergravity}
 
This model containing only one helicity $\frac{3}{2}$ gravitino and its vielbein was the first model of $D=4$ supergravity introduced in 1973. More generalised models of supergravity in higher dimensions and higher ordered supersymmetries ($\mathcal{N}$>1) are later developed during the late 1970s and early 1980s when supergravity was still a contender for the unified theory. One of the most prominent models is the $\mathcal{N}=8, D=11$ supergravity, since $\mathcal{N}=8$ is the maximally supersymmetric theory. However, the excitement around supergravity fell after the discoveries of inconsistencies within the theory \cite{cern_2019}. One major problem was anomalies that appeared during the quantisation of supergravity, for instance, looped interactions, which at the time did not have a method of cancellation \cite{nastase_2015_supergravity}. The resurgence of studies in supergravity was brought by its correspondence to superstring theory namely through AdS/CFT (Anti-de Sitter/Conformal Field Theory) correspondence. AdS space is an ansatz of Einstein's field equation in a vacuum with a negative cosmological constant. The space is maximally symmetric with the group $SO(2, D-2)$. A conformal field theory exhibits scale invariance, or more accurately, a $SU(N)$ Yang-Mill's theory that is invariant after a local Weyl transformation. A class of solutions in supergravity are called p-branes, which are the p-dimensional generalisation of particles. On the other hand, D-branes are solutions in superstring theory that couples to the endpoints of fundamental strings. In 1995, Joseph Polchinski discovered that there is a correspondence between p-branes in supergravity and D-branes in string theory. In the low-energy limit, the string theory with a large number of D-branes decouples into two non-interacting sectors --- $SU(N)$ supersymmetric Yang-Mills (SYM) on the D-branes and the free sector away from the D-branes. Meantime, at low energy, the supergravity also decouples into a low-energy free sector away from the brane and an interacting sector near the brane, which has the geometry of $AdS_{D-d} × S_d$. By identifying the free sector in string theory as just supergravity, one can infer that the interacting supergravity on $AdS_{D-d} × S_d$ is equivalent to the $SU(N)$ SYM on the $d−1$ dimensional boundary. This equivalence is known as AdS/CFT correspondence, which is the primary motivation for studying supergravity and its p-brane solutions. \cite{Hubeny_2015}
 
The project intends to explore the $N=1$, $D=11$ supergravity model and the phenomenology of its p-brane ansatz. In this review, some theoretical formalism is introduced and explained in relation to supergravity. After establishing the theoretical framework, the action of $D=11$ supergravity is explored from three aspects: the formulation of the field content in the action, its bosonic sector's relationship with string theory and the derivation of the field equations (for the bosonic sector). After applying asymmetric ansatz, these field equations can be solved, which gives rise to two p-branes solutions --- an electric 2-brane and a magnetic 5-brane. The later sections of the review shift the focus to the phenomenology of the p-branes. The black-hole liked properties of the branes, such as the mass, charge, singularity and horizons, will be discussed with regard to the BPS bound. The important method of Kaluza-Klein dimensional reduction is discussed and demonstrated in the context of branic motion. Lastly, the orbital motion of a probe brane around a parallel massive brane is explored, and numerical analysis of the circular orbit case is conducted.
 
\chapter{Theoretical Framework}
\section{Supersymmetry}
$Poincar\Acute{e}$ symmetry is governed by the ordinary Lie algebra consists of commutation relations between Lorentz generators, $M{\mu}{\nu}$, and the translation generators, $P{\mu}$,
\begin{equation}
    \begin{aligned}
    {\left[M_{\mu \nu}, M_{\rho \sigma}\right] } &=-\left(\eta_{\mu \rho} M_{\nu \sigma}+\eta_{\nu \sigma} M_{\mu \rho}-\eta_{\mu \sigma} M_{\nu \rho}-\eta_{\nu \rho} M_{\mu \sigma}\right),\\
    {\left[P_{\mu}, M_{\nu \rho}\right] } &=\left(\eta_{\mu \nu} P_{\rho}-\eta_{\mu \rho} P_{\nu}\right), \\
    {\left[P_{\mu}, P_{\nu}\right] } &=0.
    \end{aligned}
\end{equation}
Internal symmetry of the particles, such as the $SU(2)$ isospin, is also governed by the ordinary Lie algebra,
\begin{equation}
    [T_a,T_b]=f_{ab}^{c}T_c,
\end{equation}
where $f_{ab}$ is the structural constant. However, the combination of the two into a larger symmetry group with ordinary Lie algebra is forbidden by Coleman–Mandula theorem, such that $[Q_a,P_{\mu}]$ and $[Q_a,M_{\mu\nu}]$ always vanishes. 

To resolve this, graded Lie algebra is introduced, which is defined by anti-commutation instead of commutation,
\begin{equation}
    \{Q_{a},Q_{b}\}= \mbox{some generator}.
\end{equation}
In the Majorana spinor representation, the generators $Q_{a}^{i}$ ($i=1,\ldots,\mathcal{N}$ is the number of supersymmetries) have a self interacting anti-commutation relation and non-vanishing commutation relations with the rest of the generators in $Poincar\Acute{e}$ and internal symmetry as followed
\begin{equation}\label{2.4}
\begin{aligned}
\left\{Q_{a}^{i}, Q_{b}^{j}\right\}&=2\left(C \gamma^{\mu}\right)_{a b} P_{\mu} \delta^{i j}+C_{a b} U^{i j}+\left(C \gamma_{5}\right)_{a b} V^{i j},\\
[Q_{a}^{i},M_{\mu\nu}]&=\frac{1}{2}(\gamma_{\mu\nu})_{a}^{b}Q_{b}^{i},\\
    [Q_{a}^{i},P_{\mu}]&=0,\\
    [Q_{a}^{i},T_{r}]&=(V_r)_j^{i}T_r^j.
\end{aligned}
\end{equation}
The second relation, $Q_{b}^{i}=Q^{ai}C_{ab}$, implies that $Q_{a}^i$ is a Majorana spinor itself. Therefore, applying $Q_{a}^i$ to a bosonic field yields a spinor field; applying $Q_{a}^i$ to a spinor field yields a bosonic field. Consequently, the generators $Q_{a}^i$ give raise to the symmetry between bosons and fermions with some spinor parameter $\epsilon_{a}$,
\begin{equation*}
\begin{aligned}
    \delta \mbox{boson} =  \mbox{fermion}; \hspace{0.8cm} \delta \mbox{fermion} =  \mbox{boson}.
\end{aligned}
\end{equation*}
The boson and the fermion, related via the generator of supersymmetry, are called the super-partners of each other. \cite{nastase_2015_susy}

One important feature of the super-partners is that they must have the same degrees of freedom. Theories consist of fermions have an additional parity symmetry, $\psi \xrightarrow{} -\psi$, and the equation of motion is invariant under such transformation. The conserved charge associated with this symmetry is $f(\mbox{mod} 2)$ to account for pair annihilation and pair production, where $f$ is the number of fermions. The quantum number for this conserved quantity is $(-1)^f$, such that fermions have quantum number of $-1$ and bosons have quantum number of $1$. This implies the anticommunation between $(-1)^n$ and $Q_{a}^i$ is zero,
\begin{equation}\label{2.5}
\{ (-1)^f, Q_{a}^i\}= (-1)^f Q_{a}^i + Q_{a}^i (-1)^f =(-1)^f Q_{a}^i + Q_{a}^i (-1)^(f-1) = 0,
\end{equation}
because the $ Q_{a}^i$ transforms a fermion to a boson and vise versa. Assume the theory has $n$ fermionic states and $m$ bosonic states, the trace of the operator $\mbox{Tr}\left( (-1)^f \right)$ is $\sum \bra{k}(-1)^f\ket{k}$, where $k$ is either a fermionic state with eigenvalue of $-1$ or a bosonic states with eigenvalue of $+1$. Therefore, $\mbox{Tr}\left( (-1)^f \right)=m-n$.
\ref{2.5} implies $(-1)^f Q_{a}^i = - Q_{a}^i (-1)^f$ and given that $Q_{1}^1 = \mathbf{1}$, $\mbox{Tr}\left( (-1)^f \right)=0$ can be deduced:
\begin{equation}
    \begin{aligned}
        \mbox{Tr}\left( (-1)^f \right)&= \mbox{Tr}\left( Q_{1}^1 (-1)^f Q_{1}^1 \right)
        &= -\mbox{Tr}\left( (-1)^f Q_{1}^1  Q_{1}^1 \right)
        &= -\mbox{Tr}\left( (-1)^f \right).
    \end{aligned}
\end{equation}
Therefore, $m=n$, hence the fermionic states and bosonic states are equal in number for given supersymmetry. This property requires the introduction of other bosonic fields in supergravity to match the extra degree of freedom of the gravitino. \cite{nastase_2015_susy}

\section{Gravitino and Vielbein}

In supergravity, the supersymmetric transformation is $\delta_Q \mbox{field} = \partial_\mu \epsilon_a$ , which implies the minimal structure required is a gauge field $\psi_{\mu \alpha}$ with a spinor index $a$ and a spacetime index $\mu$, which is called the gravitino. One can define the creation and annihilation operator of fermions, $b$ and $b^{\dagger}$, from the supersymmetry generators, such that
\begin{equation}\label{2.7}
        \{b,b^{\dagger}\}=1\hspace{1cm}
        \{b,b\}=0\hspace{1cm}
        \{b^{\dagger},b^{\dagger}\}=1
\end{equation}
The representation space of the operators are the helicity states --- $\ket{h_0}$ and $b^{\dagger}\ket{h_0}$. Commutation relation $[b^{\dagger},M_{\mu\nu}]=\frac{1}{2}b^{\dagger}$ follows from the equation $[Q_{a}^{i},M_{\mu\nu}]=\frac{1}{2}(\gamma_{\mu\nu})_{a}^{b}Q_{b}^{i}$, thus $b^{\dagger}$ has the helicity of $\frac{1}{2}$ and $b^{\dagger}\ket{h_0} = \ket{h_0-\frac{1}{2}}$. The  helicity state of the supermultiplet is $(h_0, h_0-\frac{1}{2})\oplus( -h_0+\frac{1}{2}, -h_0)$. The graviton is a helicity 2 particle , thus the gravitino, $\psi_{\mu a}$, is helicity $\frac{3}{2}$ \cite{deser_zumino_1977}. Helicity $\frac{3}{2}$ particle such as the gravitino globally follow the Rarita–Schwinger field equation 
\begin{equation}
    \left(\epsilon^{\mu \kappa \rho \nu} \gamma_{5} \gamma_{\kappa} \partial_{\rho}-i m \sigma^{\mu \nu}\right) \psi_{\nu 
    \alpha}=0.
\end{equation}

Take the minimal structure of $\mathcal{N}=1$ supergravity as an example, there is only one spinor, the gravitino.
Supergravity action is invariant under three transformations: the local Lorentz transformation, local supersymmetric transformation and the general coordinate translation. 
$\psi_\mu$ under Lorentz transformation is simply,
\begin{equation}
    \delta_L (\lambda) \psi_{\mu} = -\frac{1}{4}\lambda^{ab}\gamma_{ab}\psi_{\mu},
\end{equation}
where $\lambda$ is the Lorentz transformation parameter and $a,b$ are tangent space indices. The gravitino transforms as a scalar under the general coordinate transformation
\begin{equation}
    \delta_G (\xi) \psi_{\mu} = \xi^{\nu} (\partial_{\nu} \psi_{\mu}) + (\partial_{\nu} \xi^{\nu}) \psi_{\mu}.
\end{equation}
The infinitesimal local supersymmetry transformation of $\psi_\mu$ is defined by the covariant derivative,
\begin{equation}
    \delta_Q (\epsilon) \psi_{\mu} = \hat{D}_{\mu}\epsilon,
\end{equation}
where the $\hat{}$ denotes for the covariant derivative without the torsionless constraint.  To realise the covariant derivative, vielbein is required to couple the spinor's tangent space to the curved manifold. The vielbein is a matrix $e_{a}^{\mu}(x)$ that have the unit length
\begin{equation}
    e_{a}{}^{\mu}e_{b}{}^{\nu}g_{\mu\nu}=\eta_{ab},
\end{equation}
where $g_{\mu\nu}$ and $\eta_{ab}$ are the metric of the real spacetime and of the tangent space, respectively. The inverse matrix to the vielbein, $e^{a}{}_{\mu}$, satisfies
\begin{equation}
    e_{a}{}^{\mu}e^{a}{}_{\nu}=\delta^{\mu}_{\nu} \hspace{1cm} e_{a}{}^{\mu}e^{b}{}_{\mu}=\delta^{b}_{a}.
\end{equation}
The vielbein allows a tensor field to be converted between tangent space and the real spacetime, i.e. $V_a = e_a{}^{\mu}V_{\mu}, V_{\mu} = e^a{}_{\mu}V_{a}$. The spinor connection, $\omega_{\mu}{}^{a}{}_{ b}(e)$, can be defined from the vielbein,
\begin{equation}
    \begin{aligned}
\omega_{\mu a b}(e) &=\frac{1}{2}\left(e_{a}{}^{v} \Omega_{\mu \nu b}-e_{b}{}^{v} \Omega_{\mu \nu a}-e_{a}{}^{\rho} e_{b}{}^{\sigma} e^{c}{}_{\mu} \Omega_{\rho \sigma c}\right) \\
\Omega_{\mu \nu a} &=\partial_{\mu} e_{\nu a}-\partial_{\nu} e_{\mu a}.
\end{aligned}
\end{equation}
With the  spinor connection, the covariant derivative of vectors in the tangent space can then be transported in the general coordinate,
\begin{equation}
    D_{\mu}V^{a}=\partial_{\mu}V^{a}+\omega_{\mu}{}^{a}{}_{ b}(e)V^{b}.
\end{equation}
On the other hand the covariant derivative of the vielbein is zero, which relates the Levi-Civita connection to the spinor connection in the relation
\begin{equation}
    \partial_{\mu} e^a{}_{\nu} + \omega_{\mu}{}^{a}{}_{b}e^b{}_{\nu} - \Gamma^{\rho}_{\mu\nu}e_{\rho}^a.
\end{equation}
From this relation, the Riemann curvature tensor can be expressed in terms of the spin connection,
\begin{equation}\label{2.14}
\begin{aligned}
    R_{\mu \nu}{ }^{a} {}_b =\partial_{\mu} \omega_{\nu}{}^{a}{}_ b-\partial_{\nu} \omega_{\mu}{ }^{a}{ }_{b}&+\omega_{\mu}{ }^{a}{ }_{c} \omega_{\nu}{}^{c}{ }_{b}-\omega_{\nu}{ }^{a}{ }_{c} \omega_{\mu}{ }^{c}{}_{b}\\
    R_{\mu \nu}{ }^{\rho}{}_\sigma &= R_{\mu \nu}{}^{a} {}_b e_a{}^{\rho} e^b{}_{\sigma}.
\end{aligned}
\end{equation}
In the form field formalism, veilbein can be written as a 1-form $e^a=e^a_{\mu}dx^{\mu}$, the connection can be derived from the veilbien 1-form $de^a+\omega^a{}_b\wedge e^b=0$, and the curvature can be written as $R^{ab}_{[2]}=d\omega^{ab}+\omega^{ac}\wedge\omega_c{}^{b}$. More generally, with the inclusion of torsion, the spin connection and the curvature is,
\begin{equation}
\begin{aligned}
    \hat{\omega}_{\mu a b} &= \omega_{\mu a b} +\frac{1}{8}(\Bar{\psi}_a \gamma_{\mu} \psi_b + \Bar{\psi}_\mu \gamma_{a} \psi_b - \Bar{\psi}_\mu \gamma_{b} \psi_a)\\
        \hat{R}_{\mu \nu}{ }^{a} {}_b &=\partial_{\mu} \hat{\omega}_{\nu}{ }^{a} {}_b-\partial_{\nu} \hat{\omega}_{\mu}{ }^{a}{ }_{b}+\hat{\omega}_{\mu}{ }^{a}{ }_{c} \hat{\omega}_{\nu}{}^{c}{ }_{b}-\hat{\omega}_{\nu}{ }^{a}{ }_{c} \omega_{\mu}{ }^{c}{}_{b}.
\end{aligned}
\end{equation}
With the vielbein, the spinor can be coupled to the manifold such that the covariant derivative is defined as followed,
\begin{equation}
    \hat{D}_{\mu}=\left( \partial_\mu + \frac{1}{4} \hat{\omega}_{\mu}{}^{ab}\gamma_{ab}\right)\psi_{\mu}. \cite{tanii_2014_1}
\end{equation}

Another function of the vielbein is to more convenient express Weyl transformation. Weyl transformation is a local recalling transformation via a scalar field $\phi(x)$. The spacetime metric can go under a local Weyl transformation,
\begin{equation}
    g_{\mu\nu} \xrightarrow{} g_{\mu\nu} '=e^{\phi(x)}g_{\mu\nu},
\end{equation}
which corresponds to a transformation of the vielbein and the spinor connection
\begin{equation}
\begin{aligned}
        e_{\mu}^{a} & \xrightarrow{} e_{\mu}^{a}=e^{\phi(x)} e_{\mu}^{a}\\
        \omega_{\mu a b}&\xrightarrow{}\omega_{\mu a b}' + 2e_{\mu[a}\partial_{b]}\phi.
\end{aligned}
\end{equation}
Using the vielbein transformation and the relation in Equation \ref{2.14}, the curvature under the Weyl transformation is 
\begin{equation}
    R^{\prime}=\mathrm{e}^{-2 \phi}\left[R-2(D-1) D^{\mu} \partial_{\mu} \phi-(D-1)(D-2) \partial_{\mu} \phi \partial^{\mu} \phi \right] \cite{stelle_2002}.
\end{equation}
One can also derive the curvature which is just the Weyl transformed metric. Both methods yield the same expression for the transformed curvature.

\section{Differential Forms}\label{sec2.3}
The bosonic sector of the supergravity action in the higher dimensions contains various anti-symmetric tensor fields of different ranks. These fields are anti-symmetric in nature because they are volume elements in different dimensions, which is better understood through the notion of differential forms.
A form of order $r$ describes the volume element and its orientation at every point on a manifold. A orientated volume element is a total anti-symmetric tensor underpinned by wedge product $\wedge$,
\begin{equation}
    dx^{\mu_1}\wedge dx^{\mu_2}\wedge \ldots 
    \wedge dx^{\mu_r}= \sum_{P}\mbox{sgn}(P)dx^{\mu_{P(1)}}\wedge dx^{\mu_{P(2)}}\wedge \ldots 
    \wedge dx^{\mu_{P(r)}},
\end{equation}
where $\mbox{sgn}(P) = \epsilon^{P(1)P(2)\ldots P(r)}$ introduces the anti-symmetry. A differential $r$-form is a covector that spans a $r$-dimensional vector space a at a point $p$ on the manifold $M$  $\Omega_p^{r}(M)$. An element $\omega$ of the vector space $\Omega_p^{r}(M)$ can be expressed as
\begin{equation}
    \omega = \frac{1}{r!} \omega_{\mu_1\mu_2\ldots\mu_r} dx^{\mu_1}\wedge dx^{\mu_2}\wedge \ldots 
    \wedge dx^{\mu_r},
\end{equation}
where $\frac{1}{r!}$ is the normalisation factor to account for the $r!$ terms encoded in the wedge product; $\omega_{\mu_1\mu2\ldots\mu_r}$ can be any tensor, but is anti-symmetrised by the wedge product, so it is effectively a total anti-symmetric tensor.

The overall manifold has $m$ dimensions, therefore a vector space spanned by $r$-forms has $\begin{pmatrix} m\\r\end{pmatrix}$ dimensions. This allows the extetior product of a $r$-form and a $q$-form to produce a $(r+q)$-form in $\Omega^{(r+q)}$
\begin{equation}
(\omega \wedge \xi)\left(V_{1}, \ldots, V_{q+r}\right) =\frac{1}{q ! r !} \sum_{P \in S_{q+r}} \operatorname{sgn}(P) \omega\left(V_{P(1)}, \ldots, V_{P(q)}\right) \xi\left(V_{P(q+1)}, \ldots, V_{P(q+r)}\right).
\end{equation}
The exterior derivative of a $r$-form maps $\Omega^r$ to $\Omega^{r+1}$ in the following way,
\begin{equation}
d \omega = \frac{1}{r!} \partial_{\nu}(\omega_{\mu_1\mu_2\ldots\mu_r})dx^{\mu_1}\wedge dx^{\mu_2}\wedge \ldots 
    \wedge dx^{\mu_r}  \wedge dx^{\nu}.
\end{equation}
In a $m=3$ manifold the action of exterior derivative on the scalar $\omega_0$ is  "grad", on the vector $\omega_1$ is "curl", on a surface $\omega_2$ is "div" and on $\omega_3$ is zero, because the vector space $\Omega^4$ can not be defined on a $m=3$ manifold. Based on the definition of the differential forms and exterior derivative, the following identities are satisfied which will be used in the derivations in later sections:
\begin{itemize}
    \item $dx^{\mu_1}\wedge dx^{\mu_2}\wedge \ldots \wedge dx^{\mu_r}=0$, if a index $\mu_i$ appears twice
    \item $dx\wedge(dy+dz)=dx\wedge dy+ dx\wedge dz$
    \item $\omega_r \wedge \omega_r = 0$, if $r$ is odd
    \item $\omega_r \wedge \xi_q = (-1)^{r+q} \xi_q \wedge \omega_r$
    \item $\omega \wedge (\xi \wedge \eta) = (\omega \wedge \xi) \wedge \eta$
    \item $d(\omega_r \wedge \xi_q) = d\omega_r \wedge \xi_q + (-1)^q \omega_r \wedge d\xi_q$
    \item $d(d\omega)=0$ for any $\omega$, this is known as the Bianchi identity. \cite{nakahara_2017}
\end{itemize}

One of the motivations to use differential form is its concise integration formalism. Because forms are already volume element one can simply integrate the volume element over a compact volume  of the subspace with the same dimensionality in an oriented manifold. Integration of a function $f$ in that subspace of the manifold is
\begin{equation}
    \int_{V_r}f\omega_r = \int_{V_r}\frac{1}{r!}f \omega_{\mu_1\mu_2\ldots\mu_r} \epsilon_{\mu_1\mu_2\ldots\mu_r} dx^{\mu_1}dx^{\mu_2} \ldots  dx^{\mu_r}.
\end{equation}
If the differential form is invariant under coordinate transformation on a $m$-dimensional manifold with a metric $g$, it is called an invariant volume element. It must span a $\Omega_m$ vector space and be in the following expression
\begin{equation}
    \omega_m \equiv \Omega_m = \sqrt{|g|}dx^1\wedge dx^2\wedge\ldots\wedge dx^m
\end{equation}
The other important notion is the Hodge dual of a differential form $*\omega_r$. The integration of $f$ on a region of the manifold is naturally the product between $f$ and each volume element
\begin{equation}
    \int_{M}f \Omega_{m} = \int_{M}f\sqrt{|g|}dx^{1}\wedge dx^{2}\wedge\ldots\wedge dx^{m} = \int_{M}f\sqrt{|g|}dx^{1} dx^{2}\ldots dx^{m},
\end{equation}
and the out come is invariant under coordinate transformation. To construct an invariant integral like the one above for a $r$-form ($r\leq m$), the notion of a Hodge dual is required. The vector space of a $r$-form, $\Omega^r$ is isomorphic to $\Omega^(m-r)$, which allows the linear map, Hodge star $* : \Omega^r \xrightarrow{} \Omega^(m-r)$,  defined by
\begin{equation}
*\left(\mathrm{~d} x^{\mu_{1}} \wedge \mathrm{d} x^{\mu_{2}} \wedge \ldots \wedge \mathrm{d} x^{\mu_{r}}\right) 
\quad=\frac{\sqrt{|g|}}{(m-r) \mid} \varepsilon^{\mu_{1} \mu_{2} \ldots \mu_{r}} {}_{v_{r+1} \ldots v_{m}} \mathrm{~d} x^{\nu_{r+1}} \wedge \ldots \wedge \mathrm{d} x^{\nu_{m}},
\end{equation}
where $\varepsilon$ here is a proper tensor whose indices are raised and lowered by the metric. The Hodge dual of a $\omega$ is thereby,
\begin{equation}
    * \omega=\frac{\sqrt{|g|}}{r !(m-r) !} \omega_{\mu_{1} \mu_{2} \ldots \mu_{r}} \varepsilon_{1}^{\mu_{1} \mu_{2} \ldots \mu_{r}} v_{r+1} \ldots v_{m} \mathrm{~d} x^{\nu_{r+1}} \wedge \ldots \wedge \mathrm{d} x^{\nu_{m}}.
\end{equation}
It can be shown $\omega_r \wedge * \xi_r$ is a invariant volume element that can be a part of supergravity action. The integral of $\omega_r \wedge * \xi_r$ is
\begin{equation}\label{2.32}
\begin{aligned}
    \int_M \omega_r \wedge * \xi_r  &= \int_M \frac{1}{r!^2} \omega_{\mu_{1} \mu_{2} \ldots \mu_{r}}  \xi_{\nu_{1} \nu_{2} \ldots \nu_{r}}  \frac{\sqrt{|g|}}{(m-r) !^2}\varepsilon^{\nu_{1} \nu_{2} \ldots \nu_{r}} {}_{\mu_{r+1} \ldots \mu_{m}}\\
     & \hspace{2.7cm} \times dx^{\mu_{1}} \wedge \ldots dx^{\mu_{r}} \wedge dx^{\mu_{r+1}} \wedge \ldots \wedge dx^{\mu_{m}}\\
     &=\frac{1}{r!}\int_M \omega_{\mu_{1} \mu_{2} \ldots \mu_{r}}\xi^{\nu_{1} \nu_{2} \ldots \nu_{r}} dx^{\mu_{1}} \wedge \ldots \wedge dx^{\mu_{m}}\\
     &=\frac{1}{r!}\int_M \omega_{\mu_{1} \mu_{2} \ldots \mu_{r}}\xi^{\nu_{1} \nu_{2} \ldots \nu_{r}} dx^{\mu_{1}} \ldots dx^{\mu_{m}}
\end{aligned}
\end{equation}
which is an invariant scalar thus an inner product. The notion of inner product implies that the Hodge dual can be interpreted as the orthogonal differential form that spans a complimentary vector space. \cite{nakahara_2017_7}

\chapter{D=11 Supergravity and the p-brane Ansatz}
\section{Supergravity and \texorpdfstring{$\sigma$}-model in String Theory}\label{sec3.1}
The extended supergravity beyond $N=3$ is known to be complicated. For instance, the maximally supersymmetric supergravity theory has $N=8$, which contains 1 graviton, 8 gravitinos, 28 vectors, 56 spinors, 35 scalar and 35 pseudo-scalar particles. A complete theory of that would have terms in the action satisfying all the transformations and containing all the interactions. 
On the other hand, $N=1$ supergravity has a relatively elegant formalism with very few field components --- 1 graviton, 1 gravitino and some vector fields. Therefore, this paper will only consider the $N=1$ supergravity model. There are two primary motivations to choose the $D=11$ supergravity model. Firstly, the highest dimension for a supersymmetric representation of string theory is $D=10$, namely the type IIA and type IIB string theories. Supergravity theories can exist up to $D=11$. The spacetime of $D=11$ supergravity theory can be decomposed into a $M_{10}\times S^1$ manifold, and subsequently dimensional reduced on a $D=10$ circle. The bosonic sector of the $D=10$ supergravity after the reduction is the same as that of the type IIA string theory. Naturally, the $p$-brane solutions in the $D=11$ supergravity also exhibit this correspondence with that of the type IIA string theory \cite{cremmer_julia_scherk_1978}. The other motivation to study $D=11$ supergravity is its simplicity in field content. Because all the fields in the $D=11$ supergravity ought to satisfy the equation of motion, we need to count the on shell d.o.f, to ensure the equal numbers of fermionic states and bosonic states. The graviton is $g_{M N}$ in the metric formalism or $e_{M}^a$ in the veilbein formalism; both methods gives $D(D+1)/2$ off-shell d.o.f. The metric, after linearisation, satisfies the Klein-Gordon equation that eliminates $D$ d.o.f, which leaves $D(D+1)/2-D=44$ d.o.f on-shell. The $D=11$ gamma matrices are $32\times32$, thus the gravitino $\psi^{\alpha}_{M}$ has $n=32$ spinor components and 11 spatial components which gives $n\times D$ d.o.f. The supersymmetry transformation $\delta \psi = D_{M}\epsilon$ constrains $D$ d.o.f, so $\psi^{\alpha}_{M}$ has $n(D-1)$ off-shell d.o.f. The on-shell d.o.f. is a spinor times the gauge field minus the non-traceless part, which boils down to $n/2*(D-3) = 128$ d.o.f. The difference between the fermionic and bosonic d.o.f is 84. An antisymmetric tensor $A_{M_1 \ldots M_r}$ has $\begin{pmatrix} D\\r\end{pmatrix}$ d.o.f. By subtracting its Maxwell gauge invariance, the off-shell d.o.f. is $\begin{pmatrix} D\\r\end{pmatrix}-\begin{pmatrix} D-1\\r-1\end{pmatrix}=\begin{pmatrix} D-1\\r\end{pmatrix}$. Again the tensor satisfies the Klein-Gordon equation that restrains on-shell d.o.f. down to $\begin{pmatrix} D-2\\r\end{pmatrix}$. A rank-3 anti-symmetric tensor $A_{M_1M_2M_3}$ has exactly 84 d.o.f. which is the only possible field of the remaining bosonic sector to offset the mismatch in d.o.f. Thereby, the three fields of the $D=11$ supergravity model have assembled --- $\psi_{M}$, $A_{M_1M_2M_3}$ and $g_{MN}$ \cite{nastase_2015_supergravity} \cite{cremmer_julia_scherk_1978}. The Lagrangian of the theory is 
\begin{equation}\label{3.1}
\begin{aligned}
\mathcal{L} =&-\frac{V}{4 K^{2}} R(\omega) 
-\frac{i V}{2} {\psi}_{M} \Gamma^{M N P} D_{N}\left(\frac{\omega+\hat{\omega}}{2}\right) \psi_{P}-\frac{V}{48} F_{M N P Q} F^{M N P Q} \\
&+\frac{K V}{192}\left(\hat{\psi}_{M} \Gamma^{M N \alpha \beta \gamma \delta} \psi_{N}+12 \hat{\psi}^{\alpha} \Gamma^{\gamma \delta} \psi^{\beta}\right)\left(F_{\alpha \beta \gamma \delta}+\hat{F}_{\alpha \beta \gamma \delta}\right) \\
&+\frac{2 K}{(144)^{2}} \epsilon^{\alpha_{1} \alpha_{2} \alpha_{3} \alpha_{4} \beta_{1} \beta_{2} \beta_{3} \beta_{4} M N P} F_{\alpha_{1} \alpha_{2} \alpha_{3} \alpha_{4}} F_{\beta_{1} \beta_{2} \beta_{3} \beta_{4}} A_{M} \cite{cremmer_julia_scherk_1978}.
\end{aligned}
\end{equation}
To focus on the bosonic sector of supergravity, we can make a consistent truncation by removing the terms with the gravitino. The remaining bosonic part can be written in  differential forms, given by
\begin{equation}\label{3.2}
    I_{11}=\int d^{11} x\left\{\sqrt{-g}\left(R-\frac{1}{48} F_{[4]}^{2}\right)\right\}-\frac{1}{6} \int F_{[4]} \wedge F_{[4]} \wedge A_{[3]}\cite{stelle_2002}.
\end{equation}
From \ref{2.32}, we can identify  $\sqrt{-g}\frac{1}{48}F_{[4]}^2=\frac{1}{2}F_{[4]} \wedge * F_{[4]}$, thus  \ref{3.2} can be transformed to,
\begin{equation}\label{3.3}
    I_{11}=\int d^{11} x\sqrt{-g}R -\frac{1}{6} \int 3 F_{[4]} \wedge * F_{[4]}+ F_{[4]} \wedge F_{[4]} \wedge A_{[3]}
\end{equation}
The equation of motion of $A_{[3]}$ can be derived by adding a small variation to the gauge field $A_{[3]}=A_{[3]}+\epsilon\delta A_{[3]}$, from which the variation of the action is
\begin{equation}
    \delta I= I(A_{[3]})-I(A_{[3]}+\delta A_{[3]})=0.
\end{equation}
The curvature term in the action does not contain $A_{[3]}$, therefore only the second part is included,
\begin{equation}\label{vary}
\begin{aligned}
     I(A_{[3]}+\delta A_{[3]})=-\frac{1}{6}\int & 3\hspace{0.1cm}d(A_{[3]}+\epsilon\delta A_{[3]})\wedge * d(A_{[3]}+\epsilon \delta A_{[3]}) +\\
     &d(A_{[3]}+\epsilon \delta A_{[3]}) \wedge d(A_{[3]}+\epsilon\delta A_{[3]}) \wedge  (A_{[3]}+\epsilon\delta A_{[3]}).
\end{aligned}
\end{equation}
After the expanding \ref{vary}, the varied action becomes,
\begin{equation}\label{3.5}
    \begin{aligned}
     I(A_{[3]}+\delta A_{[3]})=-\frac{1}{6}\int 
     & 3\hspace{0.1cm}\{dA_{[3]}\wedge * dA_{[3]}+\epsilon(2* dA_{[3]}\wedge d\delta A_{[3]})\}\\
     &dA_{[3]} \wedge dA_{[3]} \wedge A_{[3]}+\epsilon dA_{[3]}\wedge dA_{[3]}\wedge \delta A_{[3]}+\\
     &2\hspace{0.1cm}\epsilon dA_{[3]}\wedge A_{[3]}\wedge d\delta A_{[3]} + \mathcal{O}(\epsilon^2).
\end{aligned}
\end{equation}
From the exterior product rule in Section \ref{sec2.3} 
\begin{equation}\label{3.6}
    \begin{aligned}
     d(A_{[3]}\wedge A_{[3]} \wedge \delta A_{[3]})&= dA_{[3]}\wedge A_{[3]} \wedge \delta A_{[3]}-A_{[3]}\wedge A_{[3]} \wedge d\delta A_{[3]},\\
     d(* A_{[3]} \wedge d \delta A_{[3]})&=d(* d A_{[3]}) \wedge \delta A_{[3]} - * d A_{[3]} \wedge d(\delta A_{[3]}).
    \end{aligned}
\end{equation}
Substitute \ref{3.6} into \ref{3.5},
\begin{equation}
    \begin{aligned}
     I(A_{[3]}+\delta A_{[3]})=-\frac{1}{6}\int 
     &(3\hspace{0.1cm} dA_{[3]}\wedge * dA_{[3]} + dA_{[3]} \wedge dA_{[3]} \wedge A_{[3]})+\\
     &\epsilon\left\{6\hspace{0.1cm} d(* dA_{[3]}) + 3\hspace{0.1cm} dA_{[3]} \wedge dA_{[3]} \right\} \wedge \delta A_{[3]}+\\
     &d \left\{* dA_{[3]} \wedge d \delta A_{[3]}-2\hspace{0.1cm} dA_{[3]}\wedge A_{[3]}\wedge \delta A_{[3]}\right\} + \mathcal{O}(\epsilon^2).
\end{aligned}
\end{equation}
The, $d \left\{* dA_{[3]} \wedge d \delta A_{[3]}-2\hspace{0.1cm} dA_{[3]}\wedge A_{[3]}\wedge \delta A_{[3]}\right\}$, term is a total derivative, which can set to vanish on the boundary. The variation of the action becomes,
\begin{equation}
    \delta I = I(A_{[3]})-I(A_{[3]}+\delta A_{[3]})= -\int\epsilon\left\{\hspace{0.1cm} d(* dA_{[3]}) + \frac{1}{2}\hspace{0.1cm} dA_{[3]} \wedge dA_{[3]} \right\} \wedge \delta A_{[3]}.
\end{equation}
Because $\delta A_{[3]}$ is arbitrary, the equation of motion for $ A_{[3]}$ is,
\begin{equation}\label{3.9}
    d (* F_{[4]}) + \frac{1}{2}\hspace{0.1cm} F_{[4]} \wedge F_{[4]}.
\end{equation}
Immediately, we can identify two conserved quantities in this action, using the exact identity $d(dA)=0$. The first quantity is directly from the Bianchi identity $dF_{[4]}=d(d A_{[3]})=0$; the conserved quantity is a Gauss's Law integral over the boundary of a $M^5$ manifold
\begin{equation}\label{3.111}
    V=\int_{\partial\Tilde{\mathcal{M}}_5}F_{[4]}
\end{equation}
which is the magnetic charge. The second quantity is from  \ref{3.9} $d (* F_{[4]}) + \frac{1}{2} F_{[4]} \wedge F_{[4]}= d(* F_{[4]}+\frac{1}{2}A_{[3]}\wedge F_{[4]})=0$; the second conserved electric charge is 
\begin{equation}\label{3.12}
U=\int_{\partial\Tilde{\mathcal{M}}_8}* F_{[4]}+\frac{1}{2}A_{[3]}\wedge F_{[4]}.    
\end{equation}
The conserved charges constitutes the super-symmetry algebra shown in \ref{2.4}
\begin{equation}
    \{ Q, Q\}=C\left(\Gamma^A P_A+ \Gamma^{AB}U_{AB}+ \Gamma^{ABCDE}V_{ABCDE}\right)
\end{equation}
where $C$ is the charge matrix and $\Gamma$'s are the gamma matrices in the corresponding dimension. \cite{stelle_2002}

The $D=11$ supergravity action in  \ref{3.3} shows that there are no matter fields coupling. However, relativistic objects such as black holes, strings and membranes can still couple to the action.  Supergravity is the effective theory of string theory in the long-wavelength limit. To show this correlation, one can start with the action of a bosonic string. A string, moving through an ambient spacetime, spans a world-sheet with coordinate $\xi=(\sigma,\tau)$. The spacetime perceived by the string on the world-sheet is a Minkowski surface embedded in curved ambient spacetime. The embedding gives the following relation,
\begin{equation}
    ds^2= g_{M N}dx^{M}dx^{N}= \eta_{MN} \partial_ix^M \partial_jx^N d\xi^a d\xi^b,
\end{equation}
which defines the induced metric on the world sheet,
\begin{equation}
    \gamma_{ab}= g_{M N} \partial_ix^M \partial_jx^N d\xi^a d\xi^b.
\end{equation}
Therefore, the Nambu-Goto action of the string is given by
\begin{equation}
    I=\frac{1}{2\pi \alpha'}\int d\sigma d\tau \sqrt{-\mbox{det}(\gamma)}.
\end{equation}
The Nambu-Goto action above has higher ordered $\partial_ix^M$ terms. To have the action in the first order, the equation of motion of $\gamma_{ab}$ is utilised to produce the Polyakov action shown below
\begin{equation}
    I=\frac{1}{4\pi \alpha'}\int d\sigma d\tau \sqrt{\gamma} \gamma^{ij}\partial_i x^M\partial_j x^N g_{MN}.
\end{equation}
The qunatisation of string gives rise to two other massless fields, an antisymmetric tensor field $A_{MN}(x)$ and a scalar $\phi$ from the trace called the dilaton.
The action of a close string, travelling in the background "condensate" of its massless modes, takes the form
\begin{equation}
    I=\frac{1}{4\pi \alpha'}\int d\sigma d\tau \left[\sqrt{-\gamma} \gamma^{ij}\partial_i x^M\partial_j x^N g_{MN}(x)+i\epsilon^{ij}\partial_i x^M\partial_jx^N A_{MN}(x)\right]+ \frac{1}{4\pi}\int \alpha' \sqrt{-\gamma}R\phi(x). \cite{nastase_2015_string}
\end{equation}

The Weyl transformation $g_{MN}\xrightarrow{}e^{2\phi}g_{MN}$ leaves the Polyakov action invariant. Weyl invariance is an addition symmetry the string action uniquely exhibit, because the cancellation of the $e^{\phi}$
\begin{equation}
    \sqrt{-\mbox{det}(\gamma)}\gamma^{ij}\xrightarrow{}e^{2\phi}\sqrt{-\mbox{det}(\gamma)}e^{-2\phi}\gamma^{ij} = \sqrt{-\mbox{det}(\gamma)}\gamma^{ij},
\end{equation} 
only occurs when world-volume is a sheet. The string action as a quantum field theory should be locally scale invariant. However, the dilaton action $I_{dil}=\frac{1}{4\pi}\int \alpha' \sqrt{-\gamma}R\phi(x)$ is not invariant under Weyl transformation even on the classical level \cite{gibbons_horowitz_townsend_1995}. By introducing the conformal gauge $\gamma_{ij}=e^{2\sigma}\delta_{ij}$, the variation of with respect to $\sigma$ can produce a traceless energy-stress tensor, hence restore the scale invariance. Upon taking the limit $\alpha\xrightarrow{}0$ the equation of motion of the $\sigma$-model be
\begin{equation}
    \begin{aligned}
    \nabla_{M_1}&\left(e^{\alpha\phi}F^{M_1M_2...M_n}\right) =0 \hspace{0.5cm}3.28a\\
    \square \phi &=\frac{\alpha}{2n!}e^{\alpha\phi}F^2 \hspace{0.5cm}3.28b\\
    R_{MN}&= \frac{1}{2}\partial_M\phi\partial_N\phi + S_{MN} \hspace{0.5cm}3.28c,
\end{aligned}
\end{equation}
where $S_{MN}$ is the source tensor containing the matter fields. \cite{callan_friedan_martinec_perry_1985} These field equations can be derived from varying the $D=10$ string frame action
\begin{equation}
    I^{\text {string }}=\int d^{10} x \sqrt{-g^{(\mathrm{s})}} e^{-2 \phi}\left[R\left(g^{(\mathrm{s})}\right)+4 \nabla_{M} \phi \nabla^{M} \phi-\frac{1}{12} F_{M N P} F^{M N P}\right],
\end{equation}
where $F_{M NP}=3!\partial_{[M}A_{NP]}$. The derivation of the field equations is shown in next section. By applying Weyl transformation the reverse way $g^{(e)}_{MN}\xrightarrow{}e^{-2\phi}g^{(s)}_{MN}$, the action can be converted to the familiar Einstein frame,
\begin{equation}\label{3.19}
    I^{\text {Einstein }}=\int d^{10} x \sqrt{-g^{(\mathrm{e})}} \left[R\left(g^{(\mathrm{s})}\right)-\frac{1}{2} \nabla_{M} \phi \nabla^{M} \phi-\frac{1}{12} e^{- \phi}F_{M N P} F^{M N P}\right].
\end{equation}

We can return to the $D=11$ supergravity action in \ref{3.3} to establish the correlation with the string action described above. Supergravity theory can be brought to lower dimensions by Kaluza-Klein dimensional reduction shown in Chapter \ref{KK-R}. The $D=11$ metric is reduced on a ten dimensional ring with the line element
\begin{equation}
ds^2_{11}= e^{-\phi/6}ds^2_{10}+ e^{-4\phi/3}(dz^2+\mathcal{A}_Mdx^M)^2,\hspace{1cm} M=0,1,\ldots,9
\end{equation}
where $g_{zz}$  manifests as the dilaton $\phi$ and $g_{Mz}$ manifests as the Kaluza-Klein vector $\mathcal{A}_M$. The action in  \ref{3.3} reduces to type IIA supergravity in $D=10$,
\begin{equation}\label{3.21}
   \begin{aligned}
I_{\mathrm{IIA}}^{\text {Einstein }}=& \int d^{10} x \sqrt{-g^{(\mathrm{e})}}\left\{\left[R\left(g^{(\mathrm{e})}\right)-\frac{1}{2} \nabla_{M} \phi \nabla^{M} \phi-\frac{1}{12} e^{-\phi} F_{M N P} F^{M N P}\right]\right.\\
&\left.-\frac{1}{48} e^{\phi / 2} F_{M N P Q} F^{M N P Q}-\frac{1}{4} e^{3 \phi / 2} \mathcal{F}_{M N} \mathcal{F}^{M N}\right\}+\mathcal{L}_{F F A}
\end{aligned} 
\end{equation}
where $\mathcal{F}_{MN}=\partial_M \mathcal{A}_N - \partial_N \mathcal{A}_M$ is the field strength of $\mathcal{A}_M$. The NS-NS sector of the $D=10$ supergravity has the same expression as the effective string action in \ref{3.19}. We can identify the scalar from dimensional reduction to be the dilaton in the $\sigma$ model. Furthermore, in the string frame only the NS-NS sector is coupled to the scalar and is subject to Weyl transformation, which is to say the conformal invariant part of $D=10$ supergravity has the identical form as the effective string theory. As will be shown in Chapter \ref{KK-R}, a consistent truncation can be made to eliminate the contribution of R-R sector --- $F_{MNPQ}$ and $\mathcal{F}_{MN}$ terms. After the truncation, the generalisation of the action \ref{3.21} in $D$ dimension is given by
\begin{equation}\label{3.25}
    I_D= \int d^{D} x \sqrt{-g}\left[R\left(g\right)-\frac{1}{2} \nabla_{M} \phi \nabla^{M} \phi-\frac{1}{2n!} e^{-\phi} F_{[n]}^2\right],
\end{equation}
with gravity $g_{MN}$, the field strength $F_{[n]}$ and the scalar $\phi$. Note here that the Chern-Simon term $\mathcal{L}_{FFA}$ is not included. The Chern-Simon term exists in odd $D$ dimensions, for instance in the $D=11$ case. For the particular Ansatz chosen in this review, the Chern-Simon term vanishes which is a consistent truncation \cite{stelle_2002}. The dimensional reduced supergravity action in \ref{3.25} is the same as the $\sigma$-model string action in \ref{3.21}. Therefore, we can conclude that supergravity is the effective field theory of string theory in the low energy limit s $\alpha'\xrightarrow{}0$ \cite{nastase_2015_string}.

\section{p-brane Ansatz}
As discussed in the previous section, the supergravity as an effective field theory has the single charged action 
\begin{equation}\label{3.23}
    I_D= \int d^{D} x \sqrt{-g}\left[R\left(g\right)-\frac{1}{2} \nabla_{M} \phi \nabla^{M} \phi-\frac{1}{2n!} e^{\alpha\phi} F_{[n]}^2\right].
\end{equation}
The variations of action with the respect to the anti-symmetric tensor $A_{[n-1]}$ and the scalar $\phi$ are trivial, which give the equations of motion
\begin{equation}
\begin{aligned}
    \nabla_{M_1}\left(\frac{\partial \mathcal{L}}{\partial(\partial_{M_1}A_{M_2...M_n})}\right)&= \nabla_{M_1}\left(e^{\alpha\phi}F^{M_1M_2...M_n}\right) =0\\
    \nabla_{M}\left(\frac{\partial \mathcal{L}}{\partial(\partial_{M}\phi)}\right) -  \frac{\partial \mathcal{L}}{\partial\phi}&=-\square \phi + \frac{\alpha}{2n!}e^{\alpha\phi}F^2=0.
\end{aligned}
\end{equation}
The variation with the respect to the metric $g_{MN}$ is similar to Einstein-Hilbert action. We can redefine the curvature tensor by absorbing the scalar $\mathbb{R}_{MN} = R_{MN}-\frac{1}{2}\partial_M\phi\partial_N\phi$ and the varied action gives is
\begin{equation}
    \delta S_g=\int d^D x \sqrt{-g}\left[\mathbb{R}_{MN}-\frac{1}{2(n-1)!}e^{\alpha\phi}F_{M\ldots M_n} \right]\delta g^{MN}+\sqrt{-g}\left[-\frac{1}{2}g_{MN}\mathcal{L}\right]\delta g^{MN},
\end{equation}
The equation of motion is therefore
\begin{equation}
    \mathbb{R}_{MN}-\frac{1}{2}g_{MN}\mathbb{R}=\frac{1}{2(n-1)!}e^{\alpha\phi}\left[F_{MM_1\ldots M_n}F_{N}{}^{M_1\ldots M_n}-\frac{1}{2n}g_{MN}F^2\right].
\end{equation}
By substituting $\frac{1}{2}g_{MN}\mathbb{R}=\frac{1}{4n!}e^{\alpha\phi}g_{MN}F^2\frac{2n-D}{2-D}$, the equation of motion can also be written as
\begin{equation}\label{3.33}
\begin{aligned}
     \mathbb{R}_{MN}&=R_{MN}-\frac{1}{2}\partial_M\phi\partial_N\phi=S_{MN}\\
    R_{MN}&= \frac{1}{2}\partial_M\phi\partial_N\phi + S_{MN},\\
    S_{MN}&=\frac{1}{2(n-1)!}e^{\alpha\phi}\left[F_{MM_1\ldots M_n}F_{N}{}^{M_1\ldots M_n}-\frac{n-1}{n(D-2)}g_{MN}F^2\right].
\end{aligned}
\end{equation}
To summarise the fields ($g_{MN},A_{[n-1]}$ and $\phi$) of supergravity are governed by the field equations derived above 
\begin{equation}\label{3.31}
\begin{aligned}
    \nabla_{M_1}&\left(e^{\alpha\phi}F^{M_1M_2...M_n}\right) =0 \\
    \square \phi &=\frac{\alpha}{2n!}e^{\alpha\phi}F^2\\
    R_{MN}&= \frac{1}{2}\partial_M\phi\partial_N\phi + S_{MN}.
\end{aligned}
\end{equation}
The field equations are  non-linear and there is no systematic way of obtaining the solution. However, a symmetric ansatz can be adopted as an attempt to solve the field equations. The ansatz takes the form of a p-brane solution which is the $p$ dimensional generalisation of particles and strings.
The ansatz should preserve the translation symmetry in the transverse space, but some spatial dimensions is reserved for the brane to reside. The spacetime where the brane occupied is referred as the world volume. The space othogonal to the world volume, where the brane travels is the transverse space. To preserve some unbroken supersymmetry of the brane, the world volume metric of the ansatz is Minkowski. Overall, the ansatz has $(Poincar\Acute{e})_d\times SO(D-d)$ symmetry by requiring the metric to be
\begin{equation}
\begin{aligned}
    ds^2=e^{2A}dx^{\mu}dx^{\nu}\eta_{\mu\nu}+&e^{2B}dx^{m}dx^{n}\delta_{mn}\\
    \mu= 0, 1, \ldots, p \hspace{0.7cm} \mu= p+1, &\ldots, D-1 \hspace{0.7cm} r=\sqrt{y^m y_m}
\end{aligned}
\end{equation}
where the spacetime coordinates $x^M$ is decomposed into the world volume coordinates  $x^{\mu}$ and  the transverse space coordinates $y^m$ . The $d=p+1$ dimensional world volume consists of  $p$ spatial dimensions of the brane and one temporal dimension. The metric ansatz only has $r$ dependence on each component which preserves $SO(D-d)$ symmetry.  

Immediately from the metric ansatz, the Ricci tensor $R_{MN}$ in the field equation can be determined. The metric formalism and the veilbein formalism give two approaches, but arrive at the same expression. For the ease of the calculation the vielbein formalism is used,
\begin{equation}
    g_{MN} = e_{M}{}^{\underline{E}}e_{N}{}^{\underline{F}}g_{\underline{M}\underline{N}},
\end{equation}
where the real spacetime with indices $M=(\mu,m)$ are related to the tangent space with indices $\underline{M}=(\underline{\mu},\underline{m})$. For the metric ansatz, the veilbeins are identified as
\begin{equation}
 e^{\underline{\mu}} =e^Adx^{\underline{\mu}} \hspace{0.7cm}e^{\underline{m}} =e^Bdx^{\underline{m}}
\end{equation}
where the 1-form is $e^{\underline{E}}=dx^Me_{M}{}^{\underline{E}}$. The corresponding spin connections are defined by the the torsionless condition $de^{\underline{E}} + \omega^{\underline{E}}{}_{\underline{F}}\wedge e^{\underline{F}}=0$. The curvature 2-form is given by
\begin{equation}
    \Omega^{\underline{E}}{}_{\underline{F}}^{[2]}=d\omega^{\underline{E}}{}_{\underline{F}}+\omega^{\underline{E}}{}_{\underline{D}}\wedge\omega^{\underline{D}}{}_{\underline{F}}.
\end{equation}
The curvature tensor and Ricci tensor are related to the curvature 2-form as followed
\begin{equation}
\begin{aligned}
    R^{\underline{E}}{}_{\underline{F}\underline{G}\underline{H}}&=2\Omega^{\underline{E}}{}_{\underline{F}}^{[2]}(e_{\underline{G}},e_{\underline{H}})\\
    R_{\underline{F}\underline{H}}&=2\Omega^{\underline{E}}{}_{\underline{F}}^{[2]}(e_{\underline{E}},e_{\underline{H}}) \cite{misner_2017}
\end{aligned}
\end{equation}
where $e^a\wedge e^b(e_c,e_d)=\delta^{a}_{c}\delta^{b}_{d}-\delta^{a}_{d}\delta^{b}_{c}$. Therefore, the Ricci tensors for the world volume are the transverse space are
\begin{equation}
\begin{aligned}
    R_{\underline{\mu}\underline{\nu}}&=2\Omega^{\underline{\lambda}}{}_{\underline{\mu}}^{[2]}(e_{\underline{\lambda}},e_{\underline{\nu}})+2\Omega^{\underline{n}}{}_{\underline{\mu}}^{[2]}(e_{\underline{n}},e_{\underline{\nu}})\\
    \Omega^{\underline{\lambda}}{}_{\underline{\mu}}^{[2]}&=\eta_{\underline{\mu}\underline{\sigma}}(-e^{-2B}A'^2)e^{\underline{\lambda}}\wedge e^{\underline{\sigma}}\\
    \Omega^{\underline{n}}{}_{\underline{n}}^{[2]}&=\eta_{\underline{\mu}\underline{\sigma}}
    \left[e^{-B}(-\partial_{\underline{m}}B\partial_{\underline{n}}A+\partial_{\underline{m}}\partial_{\underline{n}}A)e^{\underline{\sigma}}\wedge e^{\underline{m}}+ e^{-2B}(\partial_{\underline{s}}B\partial_{\underline{s}}Ae^{\underline{n}}\wedge e^{\underline{\sigma}}-\partial_{\underline{n}}B\partial_{\underline{s}}Ae^{\underline{s}}\wedge e^{\underline{\sigma}}). \right],
\end{aligned}
\end{equation}
After the veilbein transformation the reverse way, multiplying by $e^{-2A}$ in this case, the world volume Ricci tensor with spacetime indices is
\begin{equation}\label{3.34}
    R_{\mu\nu}=-\eta_{\mu\nu}e^{-2(A-B)}\left(A''+dA'^2+\Tilde{d}A'B'+\frac{\Tilde{d}+1}{r}A'\right)
\end{equation}
where $d=d^{(el)}=n-1$ and $\tilde{d}=d^{(sol)}=D-n-2$ and prime represents $\partial/\partial_r$. Note that the $A''+(\tilde{d}+1)r^{-1}A'$ term comes from the radial Laplacian 
$\nabla^2{A}=A''+(\tilde{d}+1)r^{-1}A'$.
Similarly, the transverse space Ricci tensor is calculated to be
\begin{equation}\label{3.35}
    \begin{aligned}
R_{m n}=&-\delta_{m n}\left(B^{\prime \prime}+d A^{\prime} B^{\prime}+\tilde{d}\left(B^{\prime}\right)^{2}+\frac{(2 \alpha+1)}{r} B^{\prime}+\frac{\alpha}{r} A^{\prime}\right) \\
&-\frac{y^{m} y^{n}}{r^{2}}\left(\tilde{d} B^{\prime \prime}+d A^{\prime \prime}-2 d A^{\prime} B^{\prime}+d\left(A^{\prime}\right)^{2}-\tilde{d}\left(B^{\prime}\right)^{2}-\frac{\tilde{d}}{r} B^{\prime}-\frac{d}{r} A^{\prime}\right).
\end{aligned}
\end{equation}
More generally, for a curved manifold of the type $M_d = M_{p+1}\times B_{D-p-1}$ with metric,
\begin{equation}
     ds^2=e^{2A}dx^{\mu}dx^{\nu}\hat{g}_{\mu\nu}+e^{2B}dx^{m}dx^{n}\tilde{g}_{mn},
\end{equation}
the Ricci tensor can be obtained the same way
\begin{equation}\label{3.37}
    \begin{aligned}
R_{\mu \nu}=& \bar{R}_{\mu \nu}-e^{2(A-B)}\left(\widetilde{\nabla}^{2} A+\widetilde{g}^{m n} \partial_{m} A\left(d^{(el)} \partial_{n} A+d_{m} \partial_{m} B\right)\right) \bar{g}_{\mu \nu} \\
R_{mn}=& \widetilde{R}_{mn}-d^{(el)} \widetilde{\nabla}_{m} \widetilde{\nabla}_{n} A-d_{m} \widetilde{\nabla}_{m} \widetilde{\nabla}_{n} B+d_{m} \partial_{m} B \partial_{n} B-d^{(el)} \partial_{m} A \partial_{n} A \\
&+2 d^{(el)} \partial_{(m} A \partial_{n)} B-\left(\widetilde{\nabla}^{2} B+\widetilde{g}^{mn} \partial_{m} B\left(d^{(el)} \partial_{n} A+d_{m} \partial_{n} B\right)\right) \widetilde{g}_{mn}
\end{aligned}
\end{equation}
By substituting $A=A(r)$, $B=B(r)$, $\hat{g}_{\mu\nu=\eta_{\mu\nu}}$ and $\tilde{g}_{mn}=\delta_{mn}$ into \ref{3.37}, \ref{3.34} \ref{3.35} can be restored.

The ansatz of $F_{[n]}$ exhibits the duality which is what demonstrates the dyonic nature of the branes. In Maxwell's case, the electric charge is a 0-brane coupled to a one-dimensional world volume. The electric field is defined by the temporal evolution of the gauge potential implying the electric charge is dynamical. The magnetic field on the other hand is governed by Bianchi's identity, which is to say the magnetic charge is a topological excitation. The same applies to the duality of ansatz for the supergravity field equation. The first ansatz is directly from the gauge potential $A_{[n-1]}$, i.e. $F_{[n]}=dA_{[n-1]}$. The gauge potential $A_{[n-1]}$ is directly coupled to the $d_{el}=n-1$ dimensional world volume. It can be interpreted as an elementary or electric charge defined by Gauss's Law integral of the equation of motion. To preserve the isoptropicity and to be total anti-symmetric, the gauge potential can only take the following form
\begin{equation}\label{3.29}
    A_{\mu_1...\mu_{n-1}} = \epsilon_{\mu_1...\mu_{n-1}}e^{C(r)}, \hspace{0.7cm} \mbox{zero for other components}
\end{equation}
that only has $r$ dependence. Therefore the field strength is given by 
\begin{equation}
    F^{(el)}_{m\mu_1...\mu_{n-1}} =\epsilon_{\mu_1...\mu_{n-1}}\partial_me^{C(r)} , \hspace{0.7cm} \mbox{zero for other components}.
\end{equation}
The last term of \ref{3.23} can also be written as $\int_\frac{1}{2}e^{\alpha}\phi*F_{[n]}\wedge F_{[n]}$. Immediately, one can infer the other ansatz to have a field strength of rank $D-n$. It is possible to identify the local gauge field of $*F_{[n]}$ that couples to a $D-n-1$ world volume, but since $*F_{[n]}$ is topological in the transverse space, such gauge field cannot be defined globally. It is important to note that the action and the equation of the motion is not of the gauge potential $A_{[n-1]}$ but some anti-symmetric field strength $F_{[n]}$, which implies instead of $A_{[n-1]}$, $F_{[n]}$ is more fundamental. For the solitonic  or magnetic brane, it is more convenient to consider the field strength $F^{(mag)}_{[D-n]}$ and its dual $*F^{(mag)}_{[D-n]}=F^{(mag)}_{[n]}$, which is not defined through the gauge potential in \ref{3.29}, but through the topological form $F^{(mag)}_{[n]}=\lambda vol(S^{n})$. Written in the tensor formalism, the field strength  is 
\begin{equation}
    F^{(mag)}_{m_1m_2\ldots m_{\Tilde{n}}}=\lambda\epsilon_{m_1m_2\ldots m_{\Tilde{n}}q}\frac{y^q}{r^{\Tilde{n}+1}}
\end{equation}
with only transverse space indices.
$F^{(mag)}_{[n]}$ is does not have the isoptropicity due to the $y^q$ term, but the action is of $*=F^{(mag)}_{[n]}\wedge=F^{(mag)}_{[n]}$, in which only $y^qy_q=r$ appears. $F^{el}_{[n]}$ is a exact form which automatically satisfy the closeness condition $dF(el)_{[n]}$ or $\partial_[qF^{(el)}_{\mu_1\mu_2\ldots \mu_n]}=0$.  Unlike $F^{(el)}_{[n]}$, $F^{(mag)}_{[n]}$ is not exact, but because it is a topological form $dF^{(mag)}_{[n]}=0$; or one can to check for the closeness condition more explicitly
\begin{equation}
    \partial_qF^{(mag)}_{m_1m_2\ldots m_{n}}=r^{-(n+1)}\left[\epsilon_{m_1m_2\ldots m_{n}q} - (n+1)\epsilon_{m_1m_2\ldots m_{n}p}y^py_q/r^2 \right]
\end{equation}
upon taking the anti-symmetrisation $[qm_1m_2\ldots m_{n}]$ on the second term it cancels out with the first term and the closeness condition is satisfied. \cite{nakahara_2017_4}

The corresponding scalar also ought to satisfy the $SO(D-d)$ symmetry in the transverse space, so can only take the form $\phi(x^M)=\phi(r)$. 

Substituting the curvature ansatz, scalar ansatz and the field strength ansatz into the field equation in \ref{3.31}, it simplifies to a system of equations of $A(r),B(r),C(r)$ and $\phi(r)$,
\begin{equation}\label{3.46}
\begin{aligned}
A^{\prime \prime}+d\left(A^{\prime}\right)^{2}+\tilde{d} A^{\prime} B^{\prime}+\frac{(\tilde{d}+1)}{r} A^{\prime} &=\frac{\tilde{d}}{2(D-2)} S^{2} \\
B^{\prime \prime}+d A^{\prime} B^{\prime}+\tilde{d}\left(B^{\prime}\right)^{2}+\frac{(2 \tilde{d}+1)}{r} B^{\prime}+\frac{d}{r} A^{\prime} &=-\frac{d}{2(D-2)} S^{2} \\
\tilde{d} B^{\prime \prime}+d A^{\prime \prime}-2 d A^{\prime} B^{\prime}+d\left(A^{\prime}\right)^{2}-\tilde{d}\left(B^{\prime}\right)^{2} & 
-\frac{\tilde{d}}{r} B^{\prime}-\frac{d}{r} A^{\prime}+\frac{1}{2}\left(\phi^{\prime}\right)^{2} =\frac{1}{2} S^{2} \\
\phi^{\prime \prime}+d A^{\prime} \phi^{\prime}+\tilde{d} B^{\prime} \phi^{\prime}+\frac{(\tilde{d}+1)}{r} \phi^{\prime} &=-\frac{1}{2} \varsigma \alpha S^{2}.
\end{aligned}
\end{equation}
The first three equations are derived from \ref{3.31}c corresponding to the $\mu\nu$, $\delta_{mn}$ and $y^my^n$ sector of the Ricci tensor, respectively. The last equation  $S_{MN}$ is derived from \ref{3.31}b. Because $S_{MN}$ is proportional to $F^2\propto C'^2$, the source on the RHS of the field equations is all in terms of $S^2\propto C'^2$. The precise expression of $S$ is given by
\begin{equation}\label{3.47}
    S= \begin{cases}\left(e^{\frac{1}{2} \alpha \phi-d A+C}\right) C^{\prime} & \text { electric: } d=n-1, \varsigma=+1 \\ \lambda\left(e^{\frac{1}{2} \alpha \phi-\tilde{d} B}\right) r^{-\tilde{d}-1} & \text { magnetic: } d=D-n-1, \varsigma=-1\end{cases}.
\end{equation}
One can make a further simplication of the field equation by imposing the linearity condition,
\begin{equation}
    dA'+\tilde{d}B'=0,
\end{equation}
which is a requirement for unbroken supersymmetry. This condition bring the elimination of $B$ in the field equation
\begin{equation}\label{3.49}
\begin{aligned}
\nabla^{2} \phi &=-\frac{1}{2} \varsigma \alpha S^{2} \\
\nabla^{2} A &=\frac{\tilde{d}}{2(D-2)} S^{2} \\
d(D-2)\left(A^{\prime}\right)^{2}+\frac{1}{2} \tilde{d}\left(\phi^{\prime}\right)^{2} &=\frac{1}{2} \tilde{d} S^{2},
\end{aligned}
\end{equation}
where the two Laplacian equations are from \ref{3.46}a and \ref{3.46}d, respectively; the last equation is from combining \ref{3.46}b and \ref{3.46}c. From the two Laplacian equations, another linearisation condition can be concluded,
\begin{equation}
    \phi'=\frac{-\varsigma\alpha(D-2)}{\tilde{d}}A'.
\end{equation}
For conciseness, the constants are regrouped to form a new constant $\Delta$,
\begin{equation}
    \Delta=\alpha^2+\frac{2d\tilde{d}}{(D-2)} ,
\end{equation} 
Under this notation, \ref{3.46}c yields
\begin{equation}
    S^2=\frac{\Delta\phi'^2}{\alpha^2}.
\end{equation}
Substituting this back to \ref{3.46}a gives
\begin{equation}
    \nabla^2\phi+\frac{\varepsilon\Delta}{2\alpha}\phi'^2=  \nabla^2(e^{\frac{\varepsilon\Delta}{2\alpha}\phi})=0.
\end{equation}
The solution to this is simply the harmonic solution to the  spherical Laplacian in the transverse $(D-d)$ dimensions
\begin{equation}
e^{\frac{\varsigma\Delta}{2\alpha}\phi}=H(r)=1+\frac{k}{\tilde{d}} \hspace{0.3cm},
\end{equation}
where $k$ is the integration constant indicates the value of the potential at spatial infinity $\phi_{r\xrightarrow{}\infty}$. The only remaining variable to solve is $C(r)$ for the electric ansatz, which is obtained by substitute $S^2$ back in \ref{3.47}
\begin{equation}
    \frac{\Delta\phi'^2}{\alpha^2}=S^2=e^{\alpha\phi-2dA+2C}C'^2.
\end{equation}
Therefore, $C$ satisfies the differential equation
\begin{equation}
    \frac{\partial}{\partial r}\left(e^{C}\right)=\frac{-\sqrt{\Delta}}{a} e^{-\frac{1}{2} a \phi+d A} \phi^{\prime}.
\end{equation}
To summarise, all the variables in the field equations are determined
\begin{equation}
    \begin{aligned}
    e^{\phi}&=H(r)^{\frac{2\alpha}{\varsigma\Delta}}\\
    e^{A}&=H(r)^{{\frac{-4\tilde{d}}{\Delta(D-2)}}}+e^{A_{\infty}}\\
    e^{B}&=H(r)^{{\frac{4d}{\Delta(D-2)}}}+e^{B_{\infty}}\\
    \end{aligned}
\end{equation}
where the linearity condition implies $A, B$ and $\phi$ are differed by integration constants that again can be set to zero for simplicity.  $C(r)$ in the electric brane case is
\begin{equation}
    e^C=\frac{2}{\Delta}H^{-1}
\end{equation}
and the magnetic parameter $\lambda$ is related to the integration constant by
\begin{equation}
    k=\frac{\Delta}{2\tilde{d}}\lambda.
\end{equation}
After substituting the solutions above, the ansatz of the metric, the scalar and the field strength can be arrived
\begin{equation}\label{3.60}
\begin{aligned}
    ds^2&=H(r)^{{\frac{-4\tilde{d}}{\Delta(D-2)}}}dx^{\mu}dx^\nu\eta_{\mu\nu}+H(r)^{{\frac{4d}{\Delta(D-2)}}}dy^mdy^n\delta_{mn}\\
    e^\phi&= H^{\frac{2\alpha}{\varsigma\Delta}}\hspace{1cm}
    \begin{cases}  \text { electric: } & \varsigma=+1 \\ \text { magnetic: } & \varsigma=-1\end{cases}\\
    F^{(el)}_{m\mu_1\ldots\mu_{n-1}}&=\frac{2}{\sqrt{\Delta}}\epsilon_{\mu_1\ldots\mu_{n-1}}\partial_m(H^{-1})\\
    F^{(mag)}_{m_1\ldots m_{n}}&=\frac{2}{\sqrt{\Delta}}\epsilon_{m_1\ldots m_{n}r}\partial_r(H).
\end{aligned}
\end{equation}

In the context of the $D=11$ supergravity model, there are several properties worth mentioning. Firstly, in the electric case, 
the 4-form field strength implies 3-form gauge potential directly coupled to the world volume, thus a 2-brane.
The world volume has three dimensions  ($d=3$, $\tilde{d}=D-d-2=6$) and the transverse space has eight dimensions. 
In the magnetic case, the 4-form field strength couples to a five-dimensional transverse space, which leaves a six-dimensional world volume ($d=6$, $\tilde{d}=D-d-2=3$), thus a 5-brane.

Secondly, the scalar $\phi$ originates from the metric being dimensional reduced from $D=11$ to $D=10$.  This requires the coupling parameter $\alpha=0$ for the scalar to be consistently truncated in $D=11$, hence the value of $\Delta$ can be determined
\begin{equation}
\begin{aligned}
    0=\alpha^2&= \Delta - \frac{2d\tilde{d}}{D-2}\\
    \Delta&=4,
\end{aligned}
\end{equation}
which is the same for both the electric and magnetic cases.

With the field strength ansatz, one can validate the truncation of Chern-Simon term $\mathcal{FFA}$. In the electric case, $F^{(el)}_{m\mu_1\ldots\mu_{3}}$ has three world volume indices, but since the world volume only has three dimensions, there will be repetitive indices in the Levi-Civita symbol upon having the exterior product with itself $F\wedge F$. A similar argument can be made for the magnetic case. Therefore, the Chern-Simon term vanishes for both cases and does not contribute to the field equations. \cite{stelle_2002}

\chapter{BPS Bound and Horizons}
\section{Mass and Charge Density}
The brane solution exhibit conserved electric and magnetic charges which are denoted by $U$ and $V$ in \ref{3.111} and \ref{3.12}. In this section, the relationship between the electric/magnetic charge density and the mass density on the brane is explored in the context of BPS bound. 

From the spacetime metric of the ansatz, one can infer a brane source being place at the origin of the ($y^m=0$) transverse space. By taking the divergent integral in the transverse space, the mass density of the brane can be deduced. Suppose that the metric is asymptotically flat and of the form $g_{ij}=\psi\eta_{ij}+\mathcal{O}(r^{-1})$, where $\psi_{r\xrightarrow{}\infty}=1$. From Einstein's field equation, the curvature $R_{g_{ij}}$ defines the mass density in the spacetime. Through the scalar constrain equation, the mass of the source at spatial infinity given by an integration over the spacetime volume, or equivalently, a divergent integral at the boundary of that volume
\begin{equation}
    m=\mbox{lim}_{r\xrightarrow{}\infty}\int_{S(0,R)}g^{ij}g^{lk}(\partial_kg_{li}-\partial_ig_{lk})\sqrt{\mbox{det}g}\left.dS^i\right.
\end{equation}
where $\left.d S_{N}=\partial_{N}\right\rfloor d y^{m_1} \wedge ... \wedge d y^{m_{\Tilde{d}}}$ is a surface element on the boundary. More specifically, when the metric takes the form $g_{ij}=\eta_{ij}+\mathcal{O}(r^{-1/2})$, the mass at spatial infinity is given by the ADM expression
\begin{equation}
    \mathcal{E}=\mbox{lim}_{r\xrightarrow{}\infty}\int_{S(0,R)}(\partial_Mg_{MN}-\partial_Ng_{MM})dS^N. \cite{https://doi.org/10.48550/arxiv.gr-qc/0408083}
\end{equation}

The brane ansatz obtained in the last section is an effectively point-like source in the transverse space, whose location is defined by the world volume $\delta$-function. Different to the point mass case, however, the brane's would volume extends to infinity, and the integral over the transverse space is divergent. A more sensible quantity is the mass density of the brane (analogous to the tension of the string), which is obtained from integration over the boundary of the $\Tilde{d}+1$-dimensional bulk
\begin{equation}
    m=\int_{\partial\mathcal{M}_{(\Tilde{d}+1)}}(\partial^n g_{mn}-\partial_m g^b_b)d^{D-d-1}S^m.
\end{equation}
The $m, n$ are transverse space indices and $b$ accounts for the transverse space and the spatial part of the world volume. $d^{D-d-1}S^m=r^{\Tilde{d}}y^md\Omega^{D-d-1}$ is a surface element on the boundary of the $D-d$-dimension transverse space. The overall spacetime metric can then be linearised as $g_{MN}=\eta_{MN}+h_{MN}$, where $\eta_{MN}$ contains term up to $r^{-\Tilde{d}}$, where lower powered term vanishes at spatial infinity. From the brane ansatz in \ref{3.60}, $h_{mn}$ and $h_b^b$ can be identified
\begin{equation}
    h_{mn}=\frac{4kd}{\Delta(D-2)r^{\Tilde{d}}}\delta_{mn}, \hspace{1cm} h_b^b=\frac{8k(d+1/2\Tilde{d})}{\Delta(D-2)r^{\Tilde{d}}}.
\end{equation}
Therefore, the ADM mass density of the brane is calculated to be
\begin{equation}
    \mathcal{E}=\frac{4k\Tilde{d}\Omega_{D-d-1}}{\Delta} =\frac{2\lambda\Omega_{D-d-1}}{\sqrt{\Delta}}.
\end{equation}
The electric brane has the mass density $\mathcal{E}^{(el)}=\lambda\Omega_7$; the magnetic brane has mass density $\mathbf{E}^{(mag)}=\lambda\Omega_4$

The elementary 2-brane has a conserved charge density from the equation of the motion of its gauge potential $A_{[3]}$. The charge density is expressed as an integral of 8-dimensional transverse space volume or an integral at the 7-dimensional boundary 
\begin{equation}
    U=\int_{\partial\Tilde{\mathcal{M}}_8}* F_{[4]}+\frac{1}{2}A_{[3]}\wedge F_{[4]}.
\end{equation}
The Chern-Simon term vanishes for the ansatz, therefore does not contribute to the equation of motion and the charge density simply reduces to
\begin{equation}
\begin{aligned}
    U&=\int_{\partial\Tilde{\mathcal{M}}_8}* F^{(ele)}_{[4]}
    &=\int_{\partial\Tilde{\mathcal{M}}_8}d^7S^m F_{m012}=\lambda\Omega_7
\end{aligned}
\end{equation}
Though the the magnetic brane occupies a $d=6$ world volume, the more fundamental form field defines the magnetism is the field strength $F$ that couples to the $\Tilde{d}=5$ transverse space. The conserved magnetic charge simply emerged from the Bianchi's Identity $d{F}=0$. The magnetic charge density is again expressed as an 5-dimensional transverse space volume or an integral at the 4-dimensional boundary 
\begin{equation}
\begin{aligned}
    V&=\int_{\partial\Tilde{\mathcal{M}}_5} F^{(mag)}_{[4]}
    &=\int_{\partial\Tilde{\mathcal{M}}_4}d^7S^m \epsilon_{mnpqr} F^{npqr}=\lambda\Omega_4. \cite{stelle_2002}
\end{aligned}
\end{equation}

\section{BPS Bound}
Upon deriving the mass and charge densities for the brane ansatz, an intriguing equality between the two emerges
\begin{equation}
\begin{aligned}
        \mathcal{E}^{(el)}=U=\lambda\Omega_7\\
    \mathcal{E}^{(mag)}=V=\lambda\Omega_4.
\end{aligned}
\end{equation}
The relationship between the mass density and the charge density of a black-hole-like object, brane, in this case, is underpinned by a supersymmetric constraint --- BPS (Bogomolnyi–Prasad–Sommerfield) bound. 

The spinor, charge matrices and the Gamma matrices can be written in the 2-component form --- the spinor indice $A$ is splitted into two component ($\alpha,\dot{\alpha}$)
\begin{equation}
\begin{aligned}
    \psi^A&=
    \begin{pmatrix}
    \psi_{\alpha}\\
    \bar{\chi}_{\dot{\alpha}}
    \end{pmatrix}\\
    C_{AB}&=\begin{pmatrix}
    \epsilon^{\alpha\beta}&0\\
    0& \epsilon^{\dot{\alpha}\dot{\beta}}
    \end{pmatrix}\\
    \gamma^{\mu}&=\begin{pmatrix}
    0& \sigma^{\mu}\\
    \bar{\sigma}^{\mu}&0
    \end{pmatrix},
\end{aligned}
\end{equation}
where $\bar{\chi}^{\dot{\alpha}}=\epsilon^{\dot{\alpha}\alpha}(\chi^{\dot{\alpha}})^*$ and $(\sigma^{\mu})^{\alpha\dot{\alpha}}=\epsilon^{\alpha\beta}\epsilon^{\alpha\beta}(\sigma^{\mu})_{\beta\dot{\beta}}=(\mathbf{1},-\sigma^{\mu})^{\alpha\dot{\alpha}}$. 
With the presence of central charge, the algebra of the $\mathcal{N}$ extend supersymmetry is the anti-commutation relation shown in \ref{2.4},
\begin{equation}
    \left\{Q_{A}^{i}, Q_{B}^{j}\right\}=2\left(C \gamma^{\mu}\right)_{AB} P_{\mu} \delta^{i j}+C_{a b} V^{i j}+\left(C \gamma_{(2^{D-1}+1)}\right)_{AB} U^{i j}.
\end{equation}
where $\gamma_{(2^{D-1}+1)}=\gamma_1...\gamma_{2^{D-1}}$.
By adopting the 2-component formalism, the first anti-commutation relation to be drawn from \ref{2.4} is
\begin{equation}
\begin{aligned}
    \{ Q^i_{\alpha}, \bar{Q}_{j\dot{\alpha}} \}&=2(C\gamma^\mu)_{\alpha\dot{\alpha}} \delta^i_jP_\mu\\
    &=2(\sigma^\mu)_{\alpha\dot{\alpha}} \delta^i_jP_\mu
\end{aligned}
\end{equation}
where $\bar{Q}_{j\dot{\alpha}}$, since $C_{\alpha\dot{\alpha}}$ and $(C\gamma_{(2^{D-1}+1)})_{\alpha\dot{\alpha}}$ are diagonal term which are zero. In the rest frame, where $P^\mu=(M,0,0,0)$ this anti-commutation relation simply reduces to
\begin{equation}
    \{ Q^i_{\alpha}, \bar{Q}_{j\dot{\alpha}} \}=2M\mathbf{1}\delta^i_j=2M\begin{pmatrix}
    \mathbf{1}& 0\\
    0&\mathbf{1}
    \end{pmatrix}.
\end{equation}
From the above relation, the creation and annihilation operators can be defined
\begin{equation}
    a_{\alpha}^i=\frac{1}{\sqrt{2M}} Q_{\alpha}^i \hspace{1cm}
    a_{\alpha}^{\dagger i}=\frac{1}{\sqrt{2M}} \bar{Q}_{i\dot{\alpha}}
\end{equation}
which satisfy the the anti-commutation algebra in \ref{2.7}
\begin{equation}\label{4.155}
    \{ a_{\alpha}^i, a_{\beta}^{\dagger j}\}=\delta^{ij}\delta_{\alpha\beta} \hspace{1cm}
    \{ a_{\alpha}^i, a_{\beta}^{j}\}= \{ a_{\alpha}^{\dagger i}, a_{\beta}^{\dagger j}\}=0.
\end{equation}
The other two anti-commutation relations are
\begin{equation}\label{4.16}
    \begin{aligned}
        \{Q_\alpha^i,Q_\beta^j\}=C_{\alpha\beta} V^{i j}+\left(C \gamma_{(2^{D-1}+1)}\right)_{\alpha\beta} U^{i j}\\
        \{Q_\alpha^i,Q_\beta^j\}=C_{\dot{\alpha}\dot{\beta}} V^{i j}+\left(C \gamma_{(2^{D-1}+1)}\right)_{\dot{\alpha}\dot{\beta}} U^{i j}
    \end{aligned}
\end{equation}
since $(C\gamma^{\mu})_{\alpha\beta}$, being on the diagonal blocks, are zero. 
To demonstrate BPS bound, the $\mathcal{N}=2$ example is used here, where the central charge can be diagonalised as $Z^{ij}=(U^{ij}+V^{ij})=(U+V)\epsilon^{ij}\hspace{0.2cm}Z\in \mathbb{R}$ after the $SU(2)$ and $U(1)$ transformation of the generators. Therefore, the anti-commutation relation in \ref{4.16} is reduced to
\begin{equation}
    \begin{aligned}
        \{Q_\alpha^i,Q_\beta^j\}=2Z\epsilon_{\alpha\beta}\epsilon^{ij}\\
        \{Q_\alpha^i,Q_\beta^j\}=2Z\epsilon_{\dot{\alpha}\dot{\beta}}\epsilon^{ij}.
    \end{aligned}
\end{equation}
 The creation and annihilation operators can be defined to satisfy the condition in \ref{4.155}
\begin{equation}
\begin{aligned}
a_{\alpha}^1 &=\frac{1}{\sqrt{2}}\left[Q_{\alpha}^{1}+\epsilon_{\alpha \dot{\beta}} \bar{Q}_{2 \dot{\beta}}\right] & a_{\alpha}^{\dagger1} &=\frac{1}{\sqrt{2}}\left[\bar{Q}_{1 \dot{\alpha}}+\epsilon_{\alpha \beta} Q_{\beta}^{2}\right] \\
a_{\alpha}^2 &=\frac{1}{\sqrt{2}}\left[Q_{\alpha}^{1}-\epsilon_{\alpha \dot{\beta}} \bar{Q}_{2 \dot{\beta}}\right] & a_{\alpha}^{\dagger2} &=\frac{1}{\sqrt{2}}\left[\bar{Q}_{1 \dot{\alpha}}-\epsilon_{\alpha \beta} Q_{\beta}^{2}\right],
\end{aligned}
\end{equation}
and its algebra follows
\begin{equation}\label{4.19}
    \{a_\alpha^1,a_\beta^{\dagger1}\}=2(M-Z)\delta_{\alpha\beta} \hspace{1cm}
    \{a_\alpha^2,a_\beta^{\dagger2}\}=2(M+Z)\delta_{\alpha\beta}.
\end{equation}
If ${a,a^\dagger}=c$, then acting the creation and annihilation operators on a wavefunction
\begin{equation}
    |a\ket{\psi}|^2+|a^\dagger\ket{\psi}|^2=|\bra{\psi}{a,a^\dagger}\ket{\psi}|^2=c|\braket{\psi}{\psi}|^2
\end{equation}
is positive, hence $c$ is positive. Thus, the relationship in \ref{4.19} gives the inequality 
\begin{equation}
    M\geq|Z|
\end{equation}
known as the BPS bound. \cite{nastase_2015_string}

With the absence of the magnetic brane, the electric brane simply has the inequality  $M\geq U$, or inversely with only the magnetic brane, the inequality is $M\geq V$.  
The only spinor in the  $\mathcal{N}=1$ and $D=11$ supergravity theory is the gravitino $\psi_{C}$; the equation of motion of the gravitino yields the conserved supercharge, which is an integral over the boundary of the $D=10$ hyper-surface
\begin{equation}
    Q_{\beta}=\int_{\partial \mathcal{M}_{10}}\Gamma^{ABC}\left. \psi_{\beta C}\right.d^{10}S_{AB}= \int_{\partial \mathcal{M}_{10}}\left.\Gamma^{0bc}\right. \psi_{\beta c}d^{10}S_{b}.
\end{equation}
The supercharge being the expression above ensures the algebra $\{Q_\alpha,Q_\beta\}$ is positive. Therefore, the LHS of \ref{2.4} again obeys the BPS bound
\begin{equation}
\begin{aligned}
    \mathcal{E}\geq U & \hspace{1cm} \mbox{electric}\\
    \mathcal{E}\geq V & \hspace{1cm} \mbox{magnetic}.
\end{aligned}
\end{equation}
The calculation conducted in the previous section implies the inequality is saturated, i.e. the mass density and the electric charge density are equal for the 2-brane; the mass density and the magnetic charge density are equal for the 5-brane. The Killing spinor equation imposes three requirements on the background spinor field, one being the linearity condition ($dA'+\Tilde{dB'=0}$). The three requirements and the saturation of the BPS bound, together preserves half of the supersymmetric transformation (32 in total), while the other half are spontaneously broken. Though half broken supersymmetry is an important implication of BPS saturation, it will not be the focus of this review. Instead, the relationship between branic horizons and the BPS bound is investigated, since it will remain an integral concept in the discussion regarding the branic motion in the next chapter.

\section{Brane Horizons}
From the previous section, we have already seen the branes are underpinned by BPS bound, which is a black hole property. This section will discuss more black hole properties the brane exhibit, namely horizons and singularities, and how these properties are related to BPS bound.

Take the electric brane as an example, the metric of the electric brane ansatz is
\begin{equation}
    ds^2=\left(1+\frac{k}{r^6}\right)^{-2/3}dx^{\mu}dx^\nu\eta_{\mu\nu}+\left(1+\frac{k}{r^6}\right)^{1/3}dy^mdy^n\delta_{mn} \hspace{0.5cm} \mu=0,\ldots,3
\end{equation}
which seems to have a naked singularity at $r=0$. However, upon substituting $A'$ and $B'$, the curvature $R$ is proportional to $1/(r^6+k)^2$, thus $R$ does not diverge when $r$ tends to zero, which implies $r=0$ is not a ture singularity. By applying the coordinate transformation $r^6=\Tilde{r}^6-k$, the metric becomes 
\begin{equation}\label{4.24}
    ds^2=\left(1-\frac{k}{\Tilde{r}^6}\right)^{2/3}dx^{\mu}dx^\nu\eta_{\mu\nu}+\left(1-\frac{k}{\Tilde{r}^6}\right)^{-2}d\Tilde{r}^2+\Tilde{r}^2 d\Omega^2_7.
\end{equation}
Under the coordinate transformation, the metric is Schwartzchild like, and the singularity is covered by a horizon at $\Tilde{r}=k$. In the case of a black hole, the surface of the horizon is on the null vector of the lightcone.  Upon crossing the horizon the lightcone flips with now the temporal vector pointing at the true singularity at $r=0$. 

For an ordinary black hole, the radial component $dr$ of the metric tends to infinity at $r=0$, which is known as a space-like singularity. Therefore space-like singularity can be interpreted as an inevitable event in the future. \cite{nastase_2015_p}
The electric brane metric also has the true singularity at $\Tilde{r}=0$. But different to a black hole, the dimensionally reduced electric brane's metric tends to infinity on the temporal component $dt$, which is a time-like singularity. A time-like singularity can be interpreted as a position in space that can only be reached by a null geodesic.  Furthermore, the horizon is degenerate, such that the lightcone does not flip over upon crossing which is a consequence of the saturation of the BPS bound. \cite{carroll_2003_gr} 
\begin{figure}[htp]
    \centering
    \includegraphics[width = 0.40\textwidth]{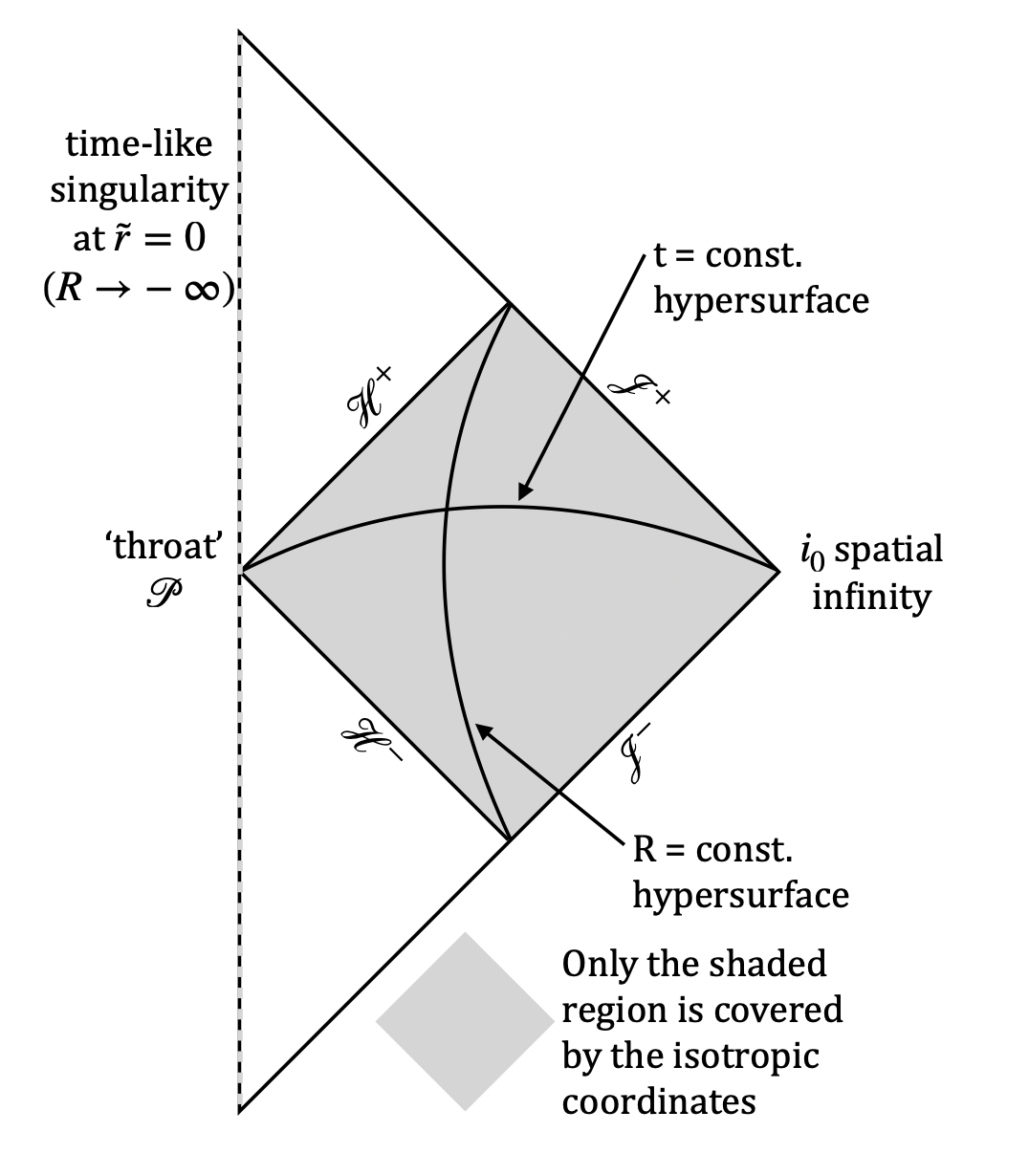}
    \caption{Carter-Penrose Diagram of the Electric Brane: $\mathcal{H}^\pm$ are the two horizons, which have the same radius and coincide; $\mathcal{J}^\pm$ and are regions of flat space region The singularity is on the temporal axis, hence is a time-like singularity.
}
    \label{fig1}
\end{figure}

Generally, the brane metric would take the form
\begin{equation}\label{4.26}
    \begin{aligned}
d s^{2}=- \frac{\Sigma_{+}}{\Sigma_{-}\left\{1-\frac{4 \tilde{d}}{\Delta(D-2)}\right\}} d t^{2}+\Sigma_{-}^{\frac{4 \tilde{d}}{\Delta(D-2)}}& d x^{i} d x^{i} 
+\frac{\Sigma_{-}^{\left\{\frac{2 a^{2}}{\Delta \tilde{d}}-1\right\}}}{\Sigma_{+}} d \tilde{r}^{2}+\tilde{r}^{2} \Sigma_{-}^{\frac{2 a^{2}}{\Delta \tilde{d}}} d \Omega_{D-d-1}^{2},\\
\Sigma_{\pm}&=1-(\frac{r_{\pm}}{\tilde{r}})^{\tilde{d}},
\end{aligned}
\end{equation}
where $r_{\pm}=M{\pm}\sqrt{M^2-U^2}$. In this metric, there are two horizons --- an inner horizon at $\tilde{r}=r_-$ and an outer horizon at $\tilde{r}=r_+$. The lightcone flips over each time when it crosses the horizon, so in the region between the two horizons the lightcone is indeed flipped like the case of a black hole. However, when the BPS bound is saturated, like in our ansatz $M=U$, the metric in \ref{4.26} reduces to the one in \ref{3.60} (which in the electric case is \ref{4.24}). In which case, the two horizons coincides $r_+=r_-$; the region between the two horizons are suppressed and the light cone does not flip over. The metric in \ref{4.26} is analogous to that of a Riessner-Nordstrom black hole. In fact, applying dimensional reduction on the electric brane, it is an Riessner-Nordstrom black hole with the additional scalar. The  extremal Riessner-Nordstrom black hole corresponds to the $M=U$ branes, and the  non-extremal Riessner-Nordstrom black hole corresponds to the $M>U$ brane or other wise known as the blackened brane. The correspondence between the electric branes and charged black holes will be more apparent in the dimensional reduction calculation in the next section, and subsequently contribute to the discussion on the orbital motion.

The metric of the magnetic brane is as followed
\begin{equation}
    ds^2=\left(1+\frac{k}{r^3}\right)^{-1/3}dx^{\mu}dx^\nu\eta_{\mu\nu}+\left(1+\frac{k}{r^6}\right)^{2/3}dy^mdy^n\delta_{mn} \hspace{0.5cm} \mu=0,\ldots,5
\end{equation}
which again seemed to have a naked singularity. By applying the coordinate transformation, $r=(\tilde{r}^3-k)^{1/3}$, the metric becomes
\begin{equation}
     ds^2=\left(1-\frac{k}{\tilde{r}^3}\right)^{1/3}dx^{\mu}dx^\nu\eta_{\mu\nu}+\left(1-\frac{k}{\tilde{r}^3}\right)^{-2} d\tilde{r}^2+\tilde{r}^2d\Omega^2_5
\end{equation}
The horizon is again located at $\tilde{r}=k$ and the true singularity is located at $\tilde{r}=0$. More interestingly, we can transform to the interpolating coordinates $r=\frac{k^{1/3}R^2}{(1-R^6)^{1/3}}$ to better understand the geometry of the magnetic ansatz, in which case the metric becomes
\begin{equation}
d s^{2}=R^{2} d x^{\mu} d x^{\nu} \eta_{\mu \nu}+k^{2 / 3}\left[\frac{4 R^{-2}}{\left(1-R^{6}\right)^{8 / 3}} d R^{2}+\frac{d \Omega_{4}^{2}}{\left(1-R^{6}\right)^{2 / 3}}\right]
\end{equation}

\begin{figure}[htp]
    \centering
    \includegraphics[width = 0.40\textwidth]{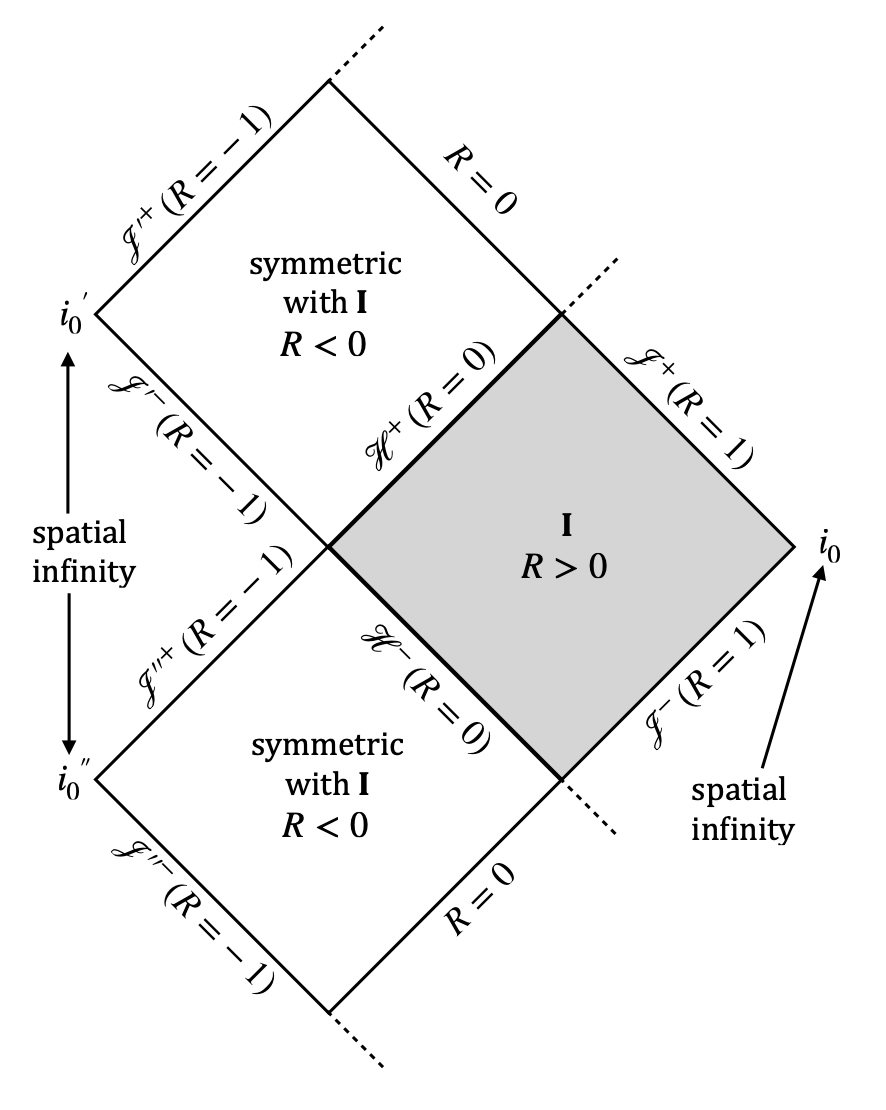}
    \caption{Carter-Penrose Diagram of the Magnetic Brane: there exists an symmetry from $R$ to $-R$. The conical singularity is the smooth extension of the horizon at $R=0$. 
}
    \label{fig2}
\end{figure}

\begin{figure}[t!]
    \centering
    \includegraphics[width = 0.60\textwidth]{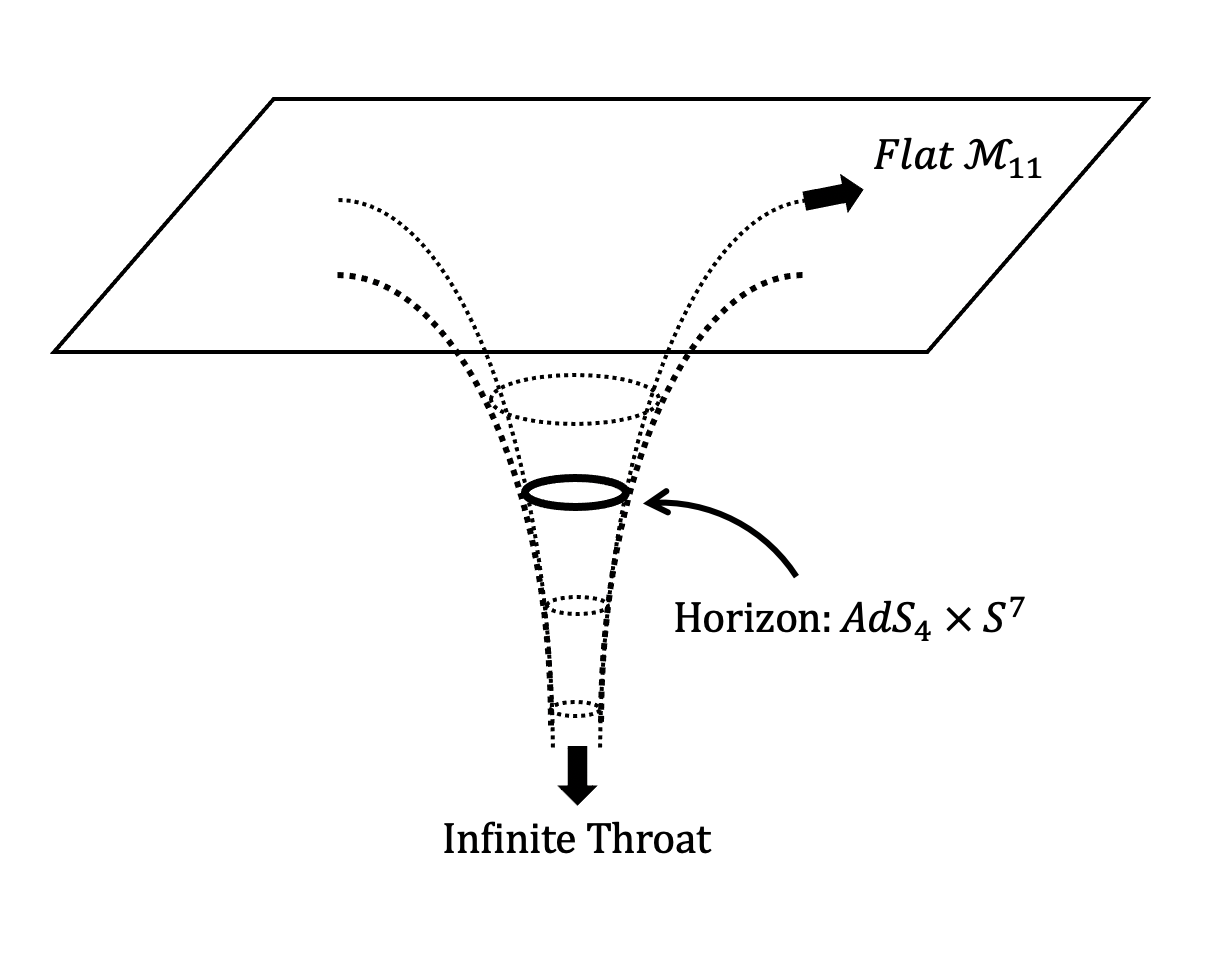}
    \caption{Interpolating Magnetic Ansatz: The spacetime interpolates between a flat $\mathcal{M}_{11}$ from afar and $AdS_4\times S^7$ on the horizon. 
}
    \label{fig3}
\end{figure}

In the interpolating soliton, the magnetic brane exhibit two interesting properties.  Firstly, the curvature at $r=R=0$ is not divergent, which implies a conical singularity (similar to the electric case). Secondly, different to the electric case the magnetic metric has an additional symmetry from $R$ to $-R$ due to its even power. Combining these two properties, one finds that the spacetime geometry of the magnetic brane is as illustrated in \ref{fig2}. The conical singularity is an infinite throat, continuously connecting the symmetric spacetime in $R$ and $-R$ \cite{stelle_2002}. The spacetime geometry of the magnetic ansatz in the interpolation coordinate is shown in Figure \ref{fig3}.  

\chapter{Dimensional Reduction}\label{KK-R}

\section{Kaluza-Klein Dimensional Reduction}
Truncation is the elimination of some field content in the theory achieved by either constraining the number of independent fields or reducing the dimension of the space-time. The former is used to obtain the bosonic sector of the action, by eliminating the gravitino terms \cite{Pons_2004}. This section focuses on reducing the spacetime dimension known as Kaluza-Klein (KK) dimensional reduction. An extra dimension was first proposed by Kaluza and Klein to unify gravity with electromagnetism. The concept of extra dimensions has since been widely used in string theory and supergravity to host bigger symmetry groups and extended objects. However, these higher dimensions, beyond the normal four dimensions, have not been observed. One postulate is that these higher dimensions are very small and compact spaces, and thus cannot be probed. The spacetime that satisfies this condition is, $\mathcal{M}_4\times K_n$, where $K_n$ is a compact space such as $S^n$ or $T^n$. Such explanation requires a method to consistently truncate the extra dimensions, so they can remain hidden in the expression of the effective lower-dimensional theory \cite{nastase_2015_kk}. A truncation is consistent if the solutions to the equation of motion of the reduced Lagrangian are the solutions to that of the original Lagrangian. \cite{Pons_2004}

As discussed before, the Type-IIA string theory exists in $D=10$, but the supergravity action is in $D+1=11$. In this context, KK dimensional reduction is used to bring the $D+1=11$ action to $D=10$ during which the dilaton emerges. To preform dimensional reduction, the coordinates need to be redefined as $x^{\hat{m}}=(x^m,z)$ where $z$ is the dimension to be reduced. With the redefinition of the coordinate, the metric $\hat{g}_{\hat{m}\hat{n}}$ can be split into three parts $g_{mn}$, $g_{mz}=g_{zm}$ and  $g_{zz}$, 
\begin{equation}\label{5.1}
\Tilde{g}_{\hat{m}\hat{n}}= e^{2a\phi}
    \begin{pmatrix}
g_{mn} + e^{2b\phi} \mathcal{A}_{m} \mathcal{A}_{n}& e^{2b\phi} \mathcal{A}_m\\
e^{2b\phi} \mathcal{A}_n& e^{2b\phi}
    \end{pmatrix},
\end{equation}
which are defined through a scalar $\phi$ and a vector $\mathcal{A}_m$. The line element in $D=11$ is described by the metric $g_{\hat{m}\hat{n}}$, which now can be rewritten as 
\begin{equation}\label{3.11}
\begin{aligned}
ds_{11}^2&=\begin{pmatrix}
dx^m&dz
\end{pmatrix}e^{2a\phi}
    \begin{pmatrix}
g_{mn} + e^{2b\phi} \mathcal{A}_{m} \mathcal{A}_{n}& e^{2b\phi} \mathcal{A}_m\\
e^{2b\phi} \mathcal{A}_n& e^{2b\phi}
    \end{pmatrix}
\begin{pmatrix}
dx^n\\
dz
\end{pmatrix}\\
&=e^{2\alpha\phi}g_{mn}dx^mdx^n+e^{2\beta\phi}(dz+\mathcal{A}_mdx^m)^2,
\end{aligned}
\end{equation}
where $\alpha=a$ and $\beta=a+b$. The scalar $\phi$ emerged from the $g_{zz}=e^{2\beta\phi}$ component of the metric corresponds to the dilaton in the $\sigma$-model. The vector $\mathcal{A}_m$ is known as the Kaluza-Klein vector. Additionally, in order to have $\sqrt{-\Tilde{g}}=\sqrt{-g}$, we should set $\beta=-(D-2)\alpha$, which also relates $\alpha$ to $g_{zz}$. \cite{duff_howe_inami_stelle_1987} The Ricci scalar $\Tilde{R}(\Tilde{g})$ can now be rewritten as a function of $\phi, R(g)$ and $ \mathcal{F}=d\mathcal{A}$
\begin{equation}
    \sqrt{-\hat{g}} R(\hat{g})=\sqrt{-g}\left(R(g)-(D-1)(D-2) \alpha^{2} \nabla_{M} \phi \nabla^{M} \phi-\frac{1}{4} e^{-2(D-1) \alpha \phi} \mathcal{F}_{[2]}{}^2\right).
\end{equation}
Conventionally, $\alpha^2$ is set to $1/[2(D-2)(D-1)]$ to normalise the kinetic term of $\phi$.
 
The other field content in \ref{3.3} is the gauge field $A_{[3]}$, which is also subjected to dimensional reduction. Generally, a $(d-1)$-form, $A_{[d-1]}$, can be written in two terms,
\begin{equation}
    A_{[d-1]}=B_{[d-1]} + B_{[d-2]} \wedge dz.
\end{equation}
The field strength $F_{[d]}=dA_{[d-1]}$ is then given by,
\begin{equation}
\begin{aligned}
        F_{[d]}&=dB_{[d-1]} + dB_{[d-2]} \wedge dz\\
        &=G_{[d]} + G_{[d-1]} \wedge dz,
\end{aligned}
\end{equation}
where $G_{[d]}=dB_{[d-1]}$ and $G_{[d-1]}=dB_{[d-2]}$. However, under this split, the Chern-Simon term $\mathcal{L}_{FFA}$, which previously vanished for our ansatz, will reappear from the dimensional reduction. Therefore, a more convenient form of field strength is adopted,
\begin{equation}
\begin{aligned}
        G_{[d-1]}&= dB_{[d-2]}\\
        G'_{[d]}=G_{[d]} - G_{[d-1]} \wedge \mathcal{A}&=dB_{[d-1]} - dB_{[d-2]} \wedge \mathcal{A},
\end{aligned}
\end{equation}
where $\mathcal{A}=\mathcal{A}_m dx^m$ is the 1-form KK vector. The field strength with now $G'_{[d]}$ and $G_{[d-1]}$ becomes,
\begin{equation}\label{5.7}
\begin{aligned}
        F_{[d]}&=dB_{[d-1]} + dB_{[d-2]} \wedge dy\\
        &=G'_{[d]} + G_{[d-1]} \wedge dz.\\\
        &=G'_{\hat{m}_1\ldots \hat{m}_d}dx^{\hat{m}}\wedge\ldots \wedge d\hat{x}^{m_d}+ G_{m_1\ldots m_{d-1}}dx^{m_1}\wedge\ldots \wedge dx^{m_{d-1}}\wedge dz
\end{aligned}
\end{equation}
The Hodge dual of $F_{[n]}$ will have an additional dilaton factor in the front to account for the countravariant indices in the Levi-Civita tensor, 
\begin{equation}\label{4.13}
\begin{aligned}
    \int F_{[d]}\wedge*F_{[d]}=\int\bigg[& \frac{1}{d!}e^{-2(d-1)\alpha\phi} G'^{\hat{m}_1\ldots \hat{m}_d}G'_{\hat{m}_1\ldots \hat{m}_d}\\
    &+\frac{1}{(d-1)!}e^{-2[(d-1)\alpha+\beta]\phi} G^{m_1\ldots m_{d-1}}G_{m_1\ldots m_{d-1}}\bigg]dx^1\wedge\ldots\wedge dx^{D}\wedge dz\\
    =\int& dx^{D}\bigg[\frac{1}{d!}e^{-2(d-1)\alpha\phi}G'_{[d]}{}^2+\frac{1}{(d-1)!}e^{2(D-d)\alpha\phi} G_{[d-1]}{}^2\bigg]
\end{aligned}
\end{equation}
Combined with the transformed curvature term, the overall action  in \ref{3.3} (without the Chern-Simon term) is dimensionally reduced to
\begin{equation}
\begin{aligned}
    I_{D}=\int d^Dx\sqrt{-g}\bigg[&R(g)- \nabla_{M} \phi \nabla^{M} \phi-\frac{1}{4} e^{-2(D-1) \alpha \phi} \mathcal{F}_{[2]}{}^2\\
    &-\frac{1}{d!}e^{-2(d-1)\alpha\phi}G'_{[d]}{}^2-\frac{1}{(d-1)!}e^{2(D-d)\alpha\phi} G_{[d-1]}{}^2.
    \bigg]
\end{aligned}
\end{equation}
By setting two of the three fields is ($\mathcal{F}_{[2]},G'_{[d]},G_{[d-1]}$), then the equation of motion of $I_D$ takes the same form as that of $I_{D-1}$ in the string frame, which is a consistent truncation. As will be shown in the following section, the ansatz used in this review has $\mathcal{F}_{[2]}=0$ and $G'_{[d]}=0$, which satisfies the condition and has the same equation of motion as the one in the dimension above. The scalar is an on-shell field that satisfies its own equation of motion, which is why the kinetic term of the scalar is kept when the theory is lifted back to $D+1=11$, which takes the expression of the single charged action in \ref{3.25}. \cite{stelle_2002}

\section{Dimensionally Reduction on Branic Motion}
The single charged action describes a static supergravity background. Having obtained the brane ansatz, one can now explore the motion of the brane in the background.
The following analysis will be on the electric 2-brane, but the procedure for dimensional reduction is similar for the magnetic 5-brane. 

The 2-brane's world volume has the coordinate $\hat{\xi}^{\hat{i}}=(t,\sigma,\rho)$ and the embedding metric $\hat{\gamma}_{\hat{i}\hat{j}}$. The motion of the 2-brane through the background is analogous to the string action, but in (2+1)-dimensional world volume. The action takes the form
\begin{equation}
S= \int \mathrm{d}^{3} \hat{\xi}\left[\frac{1}{2} \sqrt{-\hat{\gamma}} \hat{\gamma}^{\hat{i}\hat{j}} \partial_{i} \hat{x}^{\hat{m}} \partial_{j} \hat{x}^{\hat{n}} \hat{g}_{\hat{m} \hat{n}}(\hat{x})-\frac{1}{2} \sqrt{-\hat{\gamma}}\right.
\left.+\frac{1}{6} \epsilon^{\hat{i} \hat{j} \hat{k}} \partial_{\hat{i}} \hat{x}^{\hat{m}} \partial_{\hat{j}}\hat{x}^{\hat{n}} \partial_{k} \hat{x}^{\hat{p}} \hat{A}_{\hat{m} \hat{n} \hat{p}}(\hat{x})\right].
\end{equation}
Variation of  the action with the respect to $\hat{\gamma}^{\hat{i}\hat{j}}$ gives
\begin{equation}
    {\delta S}= 
    \frac{1}{2}\int \mathrm{d}^{3}\hat{\xi} \sqrt{-\hat{\gamma}}\left[\left(1 -\frac{1}{2} \hat{\gamma}_{\hat{i}\hat{j}} \hat{\gamma}^{\hat{i}\hat{j}}  \right) \left( 
    \partial_{\hat{i}}\hat{x}^{\hat{m}}\partial_{\hat{j}}\hat{x}^{\hat{n}}\hat{g}_{\hat{m} \hat{n}} \right) + \frac{1}{2}\hat{\gamma}_{\hat{i}\hat{j}}\right]\delta \hat{\gamma}^{\hat{i}\hat{j}}.
\end{equation}
Because $\hat{\gamma}_{\hat{i}\hat{j}} \hat{\gamma}^{\hat{i}\hat{j}}=3$, the variation of the action yields the following equation
\begin{equation}
    \hat{\gamma}_{\hat{i}\hat{j}} = \partial_{\hat{i}}\hat{x}^{\hat{m}}\partial_{\hat{j}}\hat{x}^{\hat{n}}\hat{g}_{\hat{m} \hat{n}}=\hat{g}_{\hat{i} \hat{j}}.
\end{equation}
This equation is implies that $\hat{\gamma}^{\hat{i}\hat{j}}$ the induced metric of $\hat{g}_{\hat{m} \hat{n}}$. If $\hat{\gamma}_{\hat{i}\hat{j}}$ satisfies its equation of motion, it can be treated as a constant while varying for $\hat{x}^{\hat{m}}$. The equation of motion for $\hat{x}^{\hat{m}}$ is obtained by substituting $\delta \hat{x}^{\hat{p}}$ in to the first part of the action, $S_1$,
\begin{equation}
\begin{aligned}
    \delta S_1&= \int d^3\hat{\xi} \left[ \sqrt{-\hat{\gamma}} \hat{\gamma}^{\hat{i}\hat{j}} \partial_{\hat{i}} \delta \hat{x}^{\hat{p}} \partial_{\hat{j}} \hat{x}^{\hat{n}} \hat{g}_{\hat{p} \hat{n}}(\hat{x})
    +\frac{1}{2} \sqrt{-\hat{\gamma}}\hat{\gamma}^{\hat{i}\hat{j}} \partial_{\hat{i}}  \hat{x}^{\hat{m}} \partial_{\hat{j}} \hat{x}^{\hat{n}} \partial_{\hat{p}}\hat{g}_{\hat{m} \hat{n}} \delta \hat{x}^{\hat{p}}
    \right]\\
    &=\int d^3\hat{\xi}  \left[ - \partial_{\hat{i}} \left(\sqrt{-\hat{\gamma}} \hat{\gamma}^{\hat{i}\hat{j}}  \partial_{\hat{j}} \hat{x}^{\hat{n}} \hat{g}_{\hat{p} \hat{n}}(\hat{x})\right) \delta \hat{x}^{\hat{p}} 
    +\frac{1}{2} \sqrt{-\hat{\gamma}}\hat{\gamma}^{\hat{i}\hat{j}} \partial_{\hat{i}}  \hat{x}^{\hat{m}} \partial_{\hat{j}} \hat{x}^{\hat{n}} \partial_{\hat{p}}\hat{g}_{\hat{m} \hat{n}} \delta \hat{x}^{\hat{p}}
    \right]\\
    &=\int d^3\hat{\xi}  \left[ - \partial_{\hat{i}} \left(\sqrt{-\hat{\gamma}} \hat{\gamma}^{\hat{i}\hat{j}}  \partial_{\hat{j}} \hat{x}^{\hat{n}}\right) \hat{g}_{\hat{p} \hat{n}} -
    \sqrt{-\hat{\gamma}} \hat{\gamma}^{\hat{i}\hat{j}} 
    \hat{g}_{\hat{p} \hat{q}} \hat{\Gamma}^{\hat{q}}_{\hat{m} \hat{n}}
    \partial_{\hat{i}}\hat{x}^{\hat{m}} \partial_{\hat{j}} \hat{x}^{\hat{n}}  \right]\delta \hat{x}^{\hat{p}}
    .
\end{aligned}
\end{equation}
Similarly the variation of the second part of the action $S_2$ is given by,
\begin{equation}
\delta S_2= \int d^3\hat{\xi} \left[\frac{1}{6} \hat{F}_{ \hat{p} \hat{m}\hat{n}  \hat{q}} \partial_{\hat{i}} \hat{x}^{\hat{m}} \partial_{\hat{j}} \hat{x}^{\hat{n}} \partial_{\hat{k}} \hat{x}^{\hat{q}} \epsilon^{\hat{i} \hat{j}\hat{k}} 
\right] \delta \hat{x}^{\hat{p}}
\end{equation}
where $\hat{F}_{ \hat{p} \hat{m}\hat{n} \hat{q}} = 4\partial_{[{\hat{p}}} \hat{A}_{\hat{m}\hat{n} \hat{q}]}$ corresponds to the differential form $F_{[4]}$.
Because $\delta \hat{x}^{\hat{p}}$ can be arbitrary, the equation of motion is given by,
\begin{equation}
\partial_{\hat{i}} \left(\sqrt{-\hat{\gamma}} \hat{\gamma}^{\hat{i}\hat{j}}  \partial_{\hat{j}} \hat{x}^{\hat{n}}\right) \hat{g}_{\hat{p} \hat{n}} +
    \sqrt{-\hat{\gamma}} \hat{\gamma}^{\hat{i}\hat{j}} 
    \hat{g}_{\hat{p} \hat{q}} \hat{\Gamma}^{\hat{q}}_{\hat{m} \hat{n}}
    \partial_{\hat{i}}\hat{x}^{\hat{m}} \partial_{\hat{j}} \hat{x}^{\hat{n}} = \frac{1}{6} \hat{F}_{ \hat{p} \hat{m}\hat{n}  \hat{q}} \partial_{\hat{i}} \hat{x}^{\hat{m}} \partial_{\hat{j}} \hat{x}^{\hat{n}} \partial_{\hat{k}} \hat{x}^{\hat{q}} \epsilon^{\hat{i} \hat{j}\hat{k}}.
\end{equation}
Multiplying the equation by $(1 / \sqrt{-\hat{\gamma}}) g^{\hat{p}\hat{s}}$ and after some relabelling gives,
\begin{equation}\label{5.16}
(1 / \sqrt{-\hat{\gamma}}) \partial_{\hat{i}}\left(\sqrt{-\hat{\gamma}} \hat{\gamma}^{\hat{i}\hat{j}} \partial_{\hat{j}} \hat{x}^{\hat{m}}\right)+\hat{\Gamma}_{\hat{n} \hat{p}}^{\hat{m}} \partial_{\hat{i}} \hat{x}^{\hat{n}} \partial_{\hat{j}} \hat{x}^{\hat{p}} \hat{\gamma}^{\hat{i}\hat{j}} 
=\frac{1}{6} \hat{F}^{\hat{m}}{ }_{\hat{n} \hat{p} \hat{q}} \partial_{\hat{i}} \hat{x}^{\hat{n}} \partial_{\hat{j}} \hat{x}^{\hat{p}} \partial_{\hat{k}} \hat{x}^{\hat{\theta}} \epsilon^{\hat{i} \hat{j}\hat{k}} / \sqrt{-\hat{\gamma}}.
\end{equation}

In the case of two branes being parallel in their world volume, dimensional reduction can be used on the metric background to remove the dependence on the spatial dimensions of the world volume. The reduced metric is similar to that of the extremal Reissner–Nordstr\"{o}m case. Furthermore, in the set-up where a probe brane orbits around a stationary big brane, the orbit of the probe brane is analogous to the orbit around a charged black hole.
Since the two electric branes are parallel, one can eliminate the two spatial dimensions in their world space by applying KK dimensional reduction twice. Without out loss of generality, $\rho$ is reduced first is by make a ten-one split of the spacetime dimensions, $\hat{x}^{\hat{m}}=\left(x^m,y\right)$, where $m=1,...,10$. More precisely, the reduced dimension is $\rho$ in the world volume, hence $y=\rho$. 
Because the two electric branes are parallel in $\rho$, the motion of the brane is independent of $\rho=y$, $\partial_{\rho}\hat{x}^{m}=0, \partial_{\rho}y=1$. Furthermore, KK dimensional reduction demands the metric, the dilaton and the gauge potential to be independent of the reduced dimension, $\partial_y\hat{g}_{\hat{m}\hat{n}} = \partial_y \hat{A}_{\hat{m}\hat{n}\hat{p}} = 0$, which is a condition satisfied by the electric brane ansatz. As obtained in Equation \ref{3.60}, the metric of the electric 2-brane is,
\begin{equation}\label{4.11}
    \hat{g}_{\hat{m}\hat{n}} =
    \begin{pmatrix}
    H^{-2/3}\hat{\eta}_{\hat{i}\hat{j}}&0\\
    0&H^{1/3}\hat{\delta}_{\hat{\mu}\hat{\nu}}
    \end{pmatrix},
\end{equation}
where $\mu,\nu$ are the transverse space indices and $H=1+\frac{k}{r^6}$ is the harmonic function for the spherical Laplacian. The line element is given by,
\begin{equation}
 d \hat{s}_{11}^{2}=H^{-2 / 3}\left(-d t^{2}+d \rho^{2}+d \sigma^{2}\right)+H^{1 / 3}\left(d r^{2}+r^{2} d \Omega_{7}^{2}\right).
\end{equation}
From the metric in  \ref{4.11}, we can obtain  $\mathcal{A}_M = 0$ (there is not diagonal term in $\hat{g}_{\hat{\mu}\hat{\nu}}$). The metric in \ref{5.1} with $\mathcal{A}_M = 0$ simply becomes,
\begin{equation}
    \hat{g}_{\hat{m}\hat{n}} =
    \begin{pmatrix}
    e^{2\alpha\phi}g_{mn}&0\\
    0&e^{2\beta\phi}
    \end{pmatrix},
\end{equation}
where one can identify $e^{2\beta\phi}=\hat{g}_{\rho\rho}=H^{-2/3}$, which gives the relation $e^{\phi}={H^{-1/(3\beta)}}$. From the previous relations, $\alpha=1/12$ and $\beta=-2/3$, we can identify $e^{\phi}=H^{-1/2}$.  The line element in $D=11$ and $D=10$ can be written as,
\begin{equation}
\begin{aligned}
 d \hat{s}_{11}^{2}&=e^{4/3\phi}\left(-d t^{2}+d \sigma^{2}+d\rho^2\right) + e^{-2/3\phi}\left(d r^{2}+r^{2} d \Omega_{7}^{2}\right)=e^{-\phi/6} ds_{10}^{2}+ e^{4\phi/3}dy^2,\\
  d s_{10}^{2}&=e^{3\phi/2}\left(-d t^{2}+d\sigma^{2}\right)+e^{-\phi/2}\left(d r^{2}+r^{2} d \Omega_{7}^{2}\right)
\end{aligned}
\end{equation}
 Therefore, the metric in $D=10$ is
\begin{equation}\label{4.15}
    g_{mn} =
    \begin{pmatrix}
    \gamma_{ij}&0\\
    0&e^{-\phi/2\phi}\delta_{\mu\nu}
    \end{pmatrix}
\end{equation}
where $\gamma_{ij} = \partial_ix^m \partial_j x^ng^{mn}$ is the induced metric in the $D=10$ with the coordinate $\xi^i=(t,\sigma)$ and the individual components are $\hat{\gamma}_{\rho\rho}=e^{4\phi/3}$ and $\hat{\gamma}_{ij}=e^{-\phi/6}\gamma_{ij}$. Consequently, the square-root determinants of the induced metrics are, 
\begin{equation}
\begin{aligned}
    \sqrt{-\hat{\gamma}}&= e^{2\phi} = H^{-2},\\
    \sqrt{-\gamma}&=e^{\frac{3}{2}\phi}=H^{-3/4},\\
    \sqrt{-\hat{\gamma}}&=e^{\frac{1}{2}\phi}\sqrt{\gamma}=H^{-1/4}\sqrt{\gamma}
\end{aligned}
\end{equation}
The equation of motion for $\hat{x}^{\hat{m}}$ in \ref{5.16} can be rewritten as 
\begin{equation}\label{5.23}
(1 / \sqrt{-\hat{\gamma}}) \partial_{\hat{i}}\left(\sqrt{-\hat{\gamma}} \hat{\gamma}^{\hat{i}\hat{j}} \partial_{\hat{j}} \hat{x}^{\hat{m}}\hat{g}_{\hat{m}\hat{p}}\right)-\frac{1}{2} \hat{\gamma}^{\hat{i}\hat{j}} \partial_{\hat{i}} \hat{x}^{\hat{m}} \partial_{\hat{j}} \hat{x}^{\hat{n}} \partial_{\hat{p}}(\hat{g}_{\hat{m}\hat{n}})
=\frac{1}{6} 4\partial_{[\hat{p}}A_{\hat{n} \hat{p} \hat{q}]} \partial_{\hat{i}} \hat{x}^{\hat{n}} \partial_{\hat{j}} \hat{x}^{\hat{p}} \partial_{\hat{k}} \hat{x}^{\hat{q}} \epsilon^{\hat{i} \hat{j}\hat{k}} / \sqrt{-\hat{\gamma}},
\end{equation}
which is a more convenient form for dimensional reduction. There is only one free index $\hat{p}$ in the equation. The $\hat{p}=y$ component of the first term in the equation is
\begin{equation}
    (1 / \sqrt{-\hat{\gamma}}) \partial_{\hat{i}}\left(\sqrt{-\hat{\gamma}} \hat{\gamma}^{\hat{i}\hat{j}} \partial_{\hat{j}} \hat{x}^{\hat{m}}\hat{g}_{\hat{m}y}\right) 
    =(1 / \sqrt{-\hat{\gamma}}) \partial_{\hat{i}}\left(\sqrt{-\hat{\gamma}} \hat{\gamma}^{\hat{i}\hat{j}} \partial_{\hat{j}} \hat{x}^{y}\hat{g}_{yy}\right)=0,
\end{equation}
since $\partial_{\hat{j}} \hat{x}^{\hat{m}}=0$.
The $\hat{p}=p$ component of the first term in the equation is
\begin{equation}
\begin{aligned}
    (1/\sqrt{-\hat{\gamma}}) \partial_{\hat{i}}\left(\sqrt{-\hat{\gamma}} \hat{\gamma}^{\hat{i}\hat{j}} \partial_{\hat{j}} \hat{x}^{\hat{m}}\hat{g}_{\hat{m}p}\right)&=(1/\sqrt{-\hat{\gamma}}) \partial_{\hat{i}}\left(\sqrt{-\hat{\gamma}} \hat{\gamma}^{\hat{i}\hat{j}} \partial_{\hat{j}} x^{m}\hat{g}_{mp}\right)\\
    &=(1/\sqrt{-\gamma}) \partial_i\left(\sqrt{-\gamma} \gamma_{ij} \partial_j x^{m}\right)g_{mp}\hspace{0.2cm}.
\end{aligned}
\end{equation}
The $\hat{p}=y$ component of the  second term in the equation is
\begin{equation}
\begin{aligned}
    \frac{1}{2} \hat{\gamma}^{\hat{i}\hat{j}} \partial_{\hat{i}} \hat{x}^{\hat{m}} \partial_{\hat{j}} \hat{x}^{\hat{n}} \partial_{y}(\hat{g}_{\hat{m}\hat{n}})=0
\end{aligned}
\end{equation}
since the metric $\hat{g}_{\hat{m}\hat{n}}$ does not depend on the reduced dimension $y$. The $\hat{p}=p$ component of the  second term in the equation is
\begin{equation}
\begin{aligned}
    \frac{1}{2} \hat{\gamma}^{\hat{i}\hat{j}} \partial_{\hat{i}} \hat{x}^{\hat{m}} \partial_{\hat{j}} \hat{x}^{\hat{n}} \partial_{p}(\hat{g}_{\hat{m}\hat{n}})=&\frac{1}{2} \gamma^{ij}\partial_ix^m\partial_jx^n  (\partial_{p}\phi)g_{mn}+\frac{1}{2} \gamma^{ij}\partial_ix^m\partial_jx^n \partial_{p}g_{mn}+\frac{1}{2} \gamma^{\rho\rho}\partial_\rho y\partial_\rho y \partial_{p}(-\phi)g_{yy}\\
    =&\frac{1}{2} \gamma^{ij}\partial_ix^m\partial_jx^n  (\partial_{p}\phi)g_{mn}+\frac{1}{2} \gamma^{ij}\partial_ix^m\partial_jx^n \partial_{p}g_{mn}-\frac{1}{2} (\partial_{p}\phi)
\end{aligned}
\end{equation}

Before the dimensional reduction on the third term of \ref{5.23}, the dimensional reduction on $A_{[3]}$ needs to be reviewed. The field strength in \ref{5.7} implies the original $ A_{[d-1]}$ is of the expression 
\begin{equation}
    A_{[d-1]}=B_{[d-1]} + B_{[d-2]} \wedge (dy+\mathcal{A}))=B_{[d-1]} + B_{[d-2]} \wedge dy.
\end{equation}
The three form of the electric ansatz is $A_{[3]}=H^{-1}dt\wedge d\sigma \wedge d\rho$, while all the other components are zero. Immediately, one can identify $\mathcal{A}=0$, $B_{[3]}=0$ and $B_{[2]}=H^{-1}dt\wedge d\sigma$. Therefore, for the electric ansatz, it satisfies the condition to have a consistent truncation ($\mathcal{F}=d\mathcal{A}=0$ and $G'_{[d]}=dB_{[d-1]}+dB_{[d-2]}\wedge\mathcal{A}=0$). 
$A_{[3]}$ can also be written as a tensor
\begin{equation}
    A_{[3]}=\frac{1}{6}\hat{\epsilon}_{\hat{i}\hat{j}\hat{k}} H^{-1} d\xi^i d\xi^j d\xi^k.
\end{equation}
More generally the tensor can be decomposed into two parts $\hat{A}_{\hat{m} \hat{n} \hat{p}}=(A_{mnp},A_{mny})$ ( $B_{[3]}=0$ and $B_{[2]}$ corresponds to $A_{mnp}=0$ and $A_{mny}$, respectively). From the reduced gauge field, the $\hat{p}=y$ RHS in \ref{5.23} is
\begin{equation}
\begin{aligned}
\frac{1}{6} \partial_{[y}A_{\hat{m} \hat{n} \hat{q}]} \partial_{\hat{i}} \hat{x}^{\hat{m}} \partial_{\hat{j}} \hat{x}^{\hat{n}} \partial_{\hat{k}} \hat{x}^{\hat{q}} \epsilon^{\hat{i} \hat{j}\hat{k}} / \sqrt{-\hat{\gamma}}=0
\end{aligned}
\end{equation}
since $A_{\hat{m} \hat{n} \hat{q}}$ has no world volume dependence.
The $\hat{p}=p$ of RHS of \ref{5.23} is
\begin{equation}
\begin{aligned}
\frac{1}{6} 4\partial_{[p}A_{mny]} \partial_{i} x^m \partial_j x^n \partial_\rho \hat{x}^y \epsilon^{ij\rho} / \sqrt{-\hat{\gamma}}&=\frac{1}{2} e^{-\phi/2} 3\partial_{[p}B_{mn]} \partial_{i} x^m \partial_j x^n  \epsilon^{ij} / \sqrt{-\gamma}\\
&=\frac{1}{2} e^{-\phi/2} F_{pmn} \partial_{i} x^m \partial_j x^n  \epsilon^{ij} / \sqrt{-\gamma}
\end{aligned}
\end{equation}
where $B_{mn}$ is the tensor from the 2-form $B_{[2]}$. 

To summarise the both LHS and RHS of \ref{5.23} vanish when the free index $\hat{p}=y$. Combining all three terms, the overall all equation of motion for $\hat{p}=p$, after reducing the dimension $y=\rho$, is 
\begin{equation}
\begin{aligned}
    \frac{1}{\sqrt{-\gamma}} \partial_i\left(\sqrt{-\gamma} \gamma^{ij} \partial_j x^{m}\right)g_{mp}-\frac{1}{2} \gamma^{ij}\partial_ix^m\partial_jx^n \partial_{p}g_{mn}&=-\frac{1}{2} \gamma^{ij}\partial_ix^m\partial_jx^n  (\partial_{p}\phi)g_{mn}-\frac{1}{2} (\partial_{p}\phi)\\
    &+\frac{1}{2}\frac{1}{\sqrt{-\gamma}} (\partial_{p}\phi)\frac{1}{2} e^{-\phi/2} F_{pmn} \partial_{i} x^m \partial_j x^n  \epsilon^{ij}.
\end{aligned}
\end{equation}

The above procedure can be repeated to further reduce on the other world volume dimension $\sigma$ that gives rise to another scalar $\varphi$. The reduced metric in $D=9$ is
\begin{equation}
\begin{aligned}
 d \hat{s}_{10}^{2}&=e^{-\varphi/2\sqrt{7}}ds_9^2+e^{-\sqrt{7}\varphi/2} dx^2\\
 d s_{9}^{2}&=e^{4\varphi/\sqrt{7}}-d t^2+e^{-2\varphi/3\sqrt{7}}\left(d r^{2}+r^{2} d \Omega_{7}^{2}\right),
\end{aligned}
\end{equation}
where $e^{\sqrt{7}\varphi/2}=H^{-3/4}$. The previous scalar $\phi$ does not change under dimensional reduction. The gauge field is reduced to $C_{[1]}=H^{-1}dt$. The similar procedure leads to the equation of motion in $D=9$ 
\begin{equation}\label{5.34}
    \frac{1}{\sqrt{-\gamma}}\frac{d}{d\tau}\left(\frac{1}{\sqrt{-\gamma}}\dot{x}^n\right)g_{np}-\frac{1}{2}\frac{1}{-\gamma} \dot{x}^n \dot{x}^p \partial_pg_{mn}=-\partial_p\left(\phi_2\right)-\frac{1}{-\gamma}\dot{x}^m\dot{x}^n\partial_{p}\left(\phi_2\right)g_{mn}-\frac{1}{\sqrt{-\gamma}} e^{-\phi_2}\partial_p C_n  \dot{x}^n,
\end{equation}
where $\phi_2=\frac{2}{\sqrt{7}}\left(\frac{1}{2}\phi+\frac{3}{\sqrt{7}}\varphi\right)$ and $e^{\phi_2}=H^{-4/7}$. By multiplying both sides by $g^{mp}$, it can also be written as
\begin{equation}
    \frac{1}{\sqrt{-\gamma}}\frac{d}{d\tau}\left(\frac{1}{\sqrt{-\gamma}}\dot{x}^n\right)+\frac{1}{-\gamma} \dot{x}^n \dot{x}^p \Gamma^m{}_{np}=-\partial^m\left(\phi_2\right)-\frac{1}{-\gamma}\dot{x}^m\dot{x}^n\partial_{n}\left(\phi_2\right)-\frac{1}{\sqrt{-\gamma}} e^{-\phi_2}\partial^m C_n  \dot{x}^n.
\end{equation}
However, this is just the equation of motion for the action 
\begin{equation}
    S_9=\int d \tau e^{\phi_2}\left[\frac{1}{2} \frac{1}{-\gamma} \dot{x}^{m} \dot{x}^{n} g_{m n}(x)+\frac{1}{2} \gamma-e^{-\phi_2} C_{m}(x) \dot{x}^{m}\right]
\end{equation}
which is of the same form as $S_{11}$, but in the string frame, which again confirms that the KK reduction on the electric brane ansatz is a consistent truncation. Thereby, both the spatial dimensions have been reduced to a point, and the action of branic motion is analogous to that around an extremal Riessner-Nordstrom black hole upto a scalar. This similarity inspires the discussion about the branic orbit in the next chapter.

\chapter{Branic Orbit}
\section{Conserved Quantity and Effective Potential}
Under dimensional reduction, the massive background brane can be reduced to a black-hole-like object. Before exploring the orbit of the probe around the background brane, it is helpful to discuss the orbits around Schwarzschild's black hole and Reissner-Nordstrom's black hole.

In principle, all the orbital can be derived from the equation of motion of $x^m$, which is a non-linear differential equation. Fortunately, because the metric, hence the Lagrangian, has no dependence on $x^m=\phi$ and $x^m=t$, the equation of motion gives the following relations
\begin{equation}
\begin{aligned}
    \frac{\partial \mathcal{L}}{\partial \phi} =\partial_{\tau} \left( \frac{\partial  \mathcal{L}}{\partial \left( \partial_{\tau} \phi \right)} \right)= \partial_{\tau} \left( \frac{\partial  \mathcal{L}}{\partial \dot{\phi}} \right)= 0,\\
    \frac{\partial \mathcal{L}}{\partial t} =\partial_{\tau} \left( \frac{\partial  \mathcal{L}}{\partial \left( \partial_{\tau} t \right)} \right)= \partial_{\tau} \left( \frac{\partial  \mathcal{L}}{\partial \dot{t}} \right)= 0.
\end{aligned}
\end{equation}
The above relations yield two conserved quantities $\frac{\partial  \mathcal{L}}{\partial \dot{t}}=-\mathcal{E}$ and $\frac{\partial  \mathcal{L}}{\partial \dot{\phi}}=L$. The conserved quantity $\mathcal{E}=\frac{e}{\mu}$ can be identified as energy density, such that, at large $r$, the energy per unit mass is just $\frac{e}{\mu}=\frac{dt}{d \tau}$. Similarly, $L = \frac{l}{\mu}$ is the angular momentum per unit mass.  In the case of the Schwartzchild black hole, the energy density and angular momentum density are 
\begin{equation}
    \begin{aligned}
        \mathcal{E}&=\left( 1 - \frac{2GM}{4}\right)\dot{t}\\
        L&=r^2\dot{\phi}.
    \end{aligned}
\end{equation}
In order to have a more intuitive understanding of the motion of the probe, we can reduce the equation of motion to the form,
\begin{equation}\label{6.7}
    \dot{r}^2+V^2=\mathcal{E}^2,
\end{equation}
where $V$ is the effective potential. Additionally, the equation of motion of $\gamma$ gives the following relation,
\begin{equation}\label{6.9}
    -\mathcal{E}^2+\dot{r}^2+\left( 1 - \frac{2GM}{r}\right)\left(\frac{L^2}{r^2}-\gamma\right)=0.
\end{equation}
By subtracting \ref{6.9} from \ref{6.7}, one obtain the expression of the effective potential purely depends the angular momentum $L$. There are three unknowns, $\dot{r}, \mathcal{E} $ and $ V$, but only two Equations \ref{6.7} and \ref{6.9}. $\dot{r}, \mathcal{E} $ and $ V$). However, $\mathcal{E}$ and $\dot{r}$ appear as $-\mathcal{E}^2 + \dot{r}^2$ in both equations, so there is effectively one unknown. Upon cancellation the effective potential is
\begin{equation}
    V(r)= -\frac{1}{2}\gamma +\gamma\frac{GM}{r}+\frac{L^2}{2r^2}-\frac{GML^2}{r^3}. \cite{carroll_2003_sc}
\end{equation}
However, for a Reissner-Nordstrom black hole, the Lagrangian contains an extra term, $\kappa A_a x^a$, which results in an additional term in $\mathcal{E}$ as well,
\begin{equation}
    \mathcal{E}=\frac{\Delta}{r^2}\dot{t}+\frac{\kappa Q}{r}.
\end{equation}
The equation of motion of $\gamma$ is now,
\begin{equation}
    \left(\mathcal{E}^2-\frac{\kappa Q}{r}\right)^2 - \dot{r}^2=\frac{\Delta}{r^2}\left(\frac{L^2}{r^2}-\gamma\right),
\end{equation}
where the terms containing $\mathcal{E}$ and $\dot{r}$ are no longer uniform with \ref{6.7}. The effective potential still be derived, but will have $\mathcal{E}$ dependence. \cite{pugliese_quevedo_ruffini_2011}

In the case of one small brane (a probe brane) orbiting around a big brane (or source), the big brane can be interpreted as the background that shapes the spacetime geometry with the metric $\hat{g}_{\hat{m}\hat{n}}$ and mediates the electric gauge potential $A_{[3]}$.  The assumption is that the probe's mass is insignificant to affect the spacetime metric, similar to a small mass compared with a black hole. When the probe is parallel to the source, one can use the dimensional reduction method described in the previous chapter to eliminate the spatial dimensions in the world volume ($\sigma$ and $\rho$) and leave the branes as point-like objects in the transverse space. The action of the probe travelling in the background is given by 
\begin{equation}\label{6.8}
 S=\int d \tau \hspace{0.1cm} e^{\frac{2 \phi_{2}(x)}{\sqrt{7}}}\left[\frac{1}{2} \sqrt{-\gamma}^{-1} \dot{x}^{m} \dot{x}^{n} g_{m n}(x)-\frac{1}{2} \sqrt{-\gamma}-\left. \epsilon Q \right. e^{-\frac{2 \phi_{2}}{\sqrt{7}}}  C_{m}(x) \dot{x}^{m}\right],
\end{equation}
where $Q=U/M$ and $\epsilon=m/\mu$ are the mass to charge ratio of the background brane and the probe, respectively; $x^m=(t, r, \theta, \theta_2,...,\theta_7)$ is the path in the nine dimensional spacetime; $e^{2\phi_2/\sqrt{7}} = H^{-4/7}$ and $C_t = H^{-1}$ while other components of $C_m$ are zero. 

In the radial coordinates, the line element is given by
\begin{equation}
    ds^2=- H^{-6/7}dt^2+H^{1/7} \left (dr^2 + r^2(d\theta^2+(d\theta_2^2+\sin^2{\theta_2}(...+\sin^2{\theta_6}d\theta_7^2) \right).
\end{equation}
One can set the motion on the probe on a seven dimensional equatorial plane, i.e. setting $\theta_1,...,\theta_7=0$, hence the line element becomes,
\begin{equation}
    ds^2=- H^{-6/7}dt^2+H^{1/7} dr^2 + H^{1/7} r^2d\theta^2.
\end{equation}
The Lagrangian has no dependence on $\theta_2, ...,\theta_7$ and their derivatives $\dot{\theta}_2,...,\dot{\theta}_7$, therefore the $x^m=\theta_{2},\ldots,\theta_{7}$ part of  \ref{6.8} is zero. Furthermore, the Lagrangian is dependent on $\dot{\theta}$ and $\dot{t}$, but is independent on $\theta$ and $t$ themselves (the metric $g_{mn}$, the scalar $\phi_2(r)$ and the gauge potential $C_m(r)$ are all independent of $t$ and $\theta$), thus the equation of motion gives the following relations, which implies that there exists two conserved quantities $\frac{\partial  \mathcal{L}}{\partial \dot{t}}$ and $\frac{\partial  \mathcal{L}}{\partial \dot{\theta}}$. The conserved charge for $t$ is
\begin{equation}
\begin{aligned}
    \frac{\partial  \mathcal{L}}{\partial \dot{t}}&=
    e^{\frac{2\phi_2}{\sqrt{7}}}\left( \dot{x}^m g_{tm}+e^{-\frac{2\phi_2}{\sqrt{7}}} \left. C_t \right. \right)\\
    &=-H^{-10/7}\dot{t}-\epsilon Q H^{-1} = -\mathcal{E} ;
\end{aligned}
\end{equation}
the conserved charge for $\theta$ is
\begin{equation}
\begin{aligned}
    \frac{\partial  \mathcal{L}}{\partial \dot{\theta}}&=
    e^{\frac{2\phi_2}{\sqrt{7}}}\left( \dot{x^m}g_{\theta m}+ \left. \epsilon Q \right. e^{-\frac{2\phi_2}{\sqrt{7}}} C_\theta \right)\\
    &=H^{-3/7}r^2\dot{\theta} = L.
\end{aligned}
\end{equation}
The equation of motion of $\gamma$ in the action \ref{6.8}
\begin{equation}\label{6.13}
\begin{aligned}
    (\mathcal{E}-\epsilon Q H^{-1})^2-H^{15/7}\dot{r}^2=H^{2}\left[\frac{HL^2}{r^2}-\gamma\right]
\end{aligned}
\end{equation}
The action in \ref{6.8} is similar to the Reissner-Nordstrom, where the gauge potential is present. Therefore, the effective potential can not be simply obtained from the two conserved quantities $\mathcal{E}$ and $l$.

\section{Circular Orbit}
\begin{figure}[htp]
    \centering
    \includegraphics[width = 0.60\textwidth]{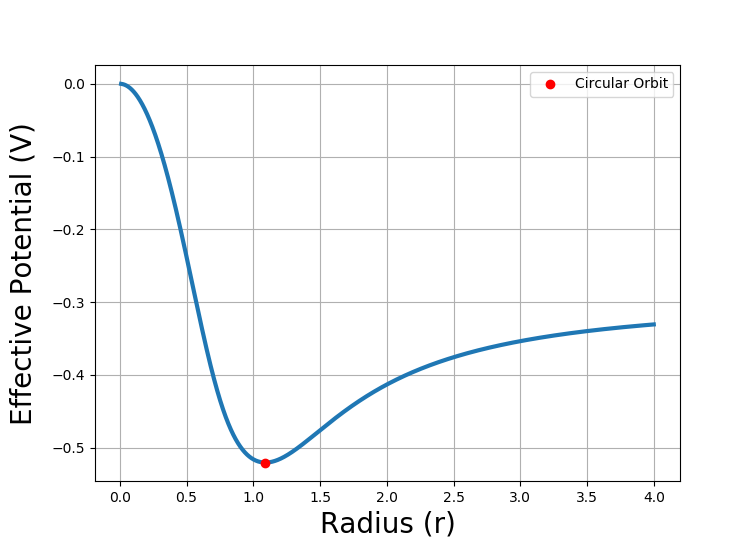}
    \caption{Relationship between the radius and effective potential of the circular orbit}
    \label{fig4}
\end{figure}

A method to simplify \ref{6.13} to obtain the effective potential is by setting $\dot{r}=0$, in which case the orbits are circular. The effective potential is therefore equal to $\mathcal{E}$
\begin{equation}
    V_{\pm}=\mathcal{E}_{\pm}=H^{-1}\left(\epsilon Q \pm  \sqrt{\frac{HL^2}{r^2}-\gamma}\right).
\end{equation}
Substituting the solution of the harmonic function $H=1+\frac{k}{r^{\Tilde{d}}}=1+\frac{k}{r^6}$ and the affine parameter $\gamma=-1$, the for expression of the effective potential is
\begin{equation}\label{6.15}
    V_{\pm}=\mathcal{E}_{\pm}=\left(1+\frac{k}{r^6}\right)^{-1}\left(\epsilon Q \pm  \sqrt{\left(1+\frac{k}{r^6}\right)\frac{L^2}{r^2}+1}\right).
\end{equation}
There are two solutions of the potential, and at $r\xrightarrow{}\infty$ they converge to $\epsilon Q \pm 1$.  By imposing the boundary condition --- in the extremal case ($\epsilon , Q =1$), the potential at $r\xrightarrow{}\infty$ vanishes --- the $V_+$ solution can be omitted; here on, $V_-$ will just be referred as $V$. Figure \ref{fig4} shows an example of the effective potential. The limit of potential at $r=0$ is $V_{r\xrightarrow{}0}=0$ instead, which again implies there is no true singularity at $r=0$. 

However, unlike the the functionality of effective potential in the Schwartzchild case, the only stable orbit can be derived the this effective potential is the circular orbit $\dot{r}=0$, i.e. the minima of the the curve, which can be solved by
\begin{equation}\label{6.16}
 \begin{aligned}
    \frac{dV}{dr}=L^2r^{12}+\left(-6\epsilon Q k \sqrt{\left(\frac{k}{r^6}+1\right)\frac{L^2}{r^2}+1}\right)r^8-L^2kr^6-2L^2k^2=0.
 \end{aligned}
\end{equation}
The location of the minima depends on the mass-charge ratio of the probe $\epsilon$, mass-charge ratio the background $Q$ and the angular momentum $L$.
\begin{figure}[htp]
    \centering
    \includegraphics[width = 0.65\textwidth]{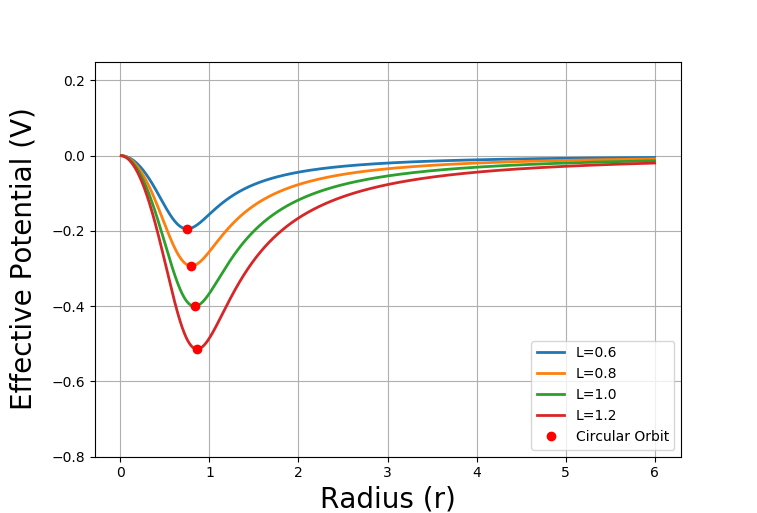}
    \caption{Effective potential under different angular momentum}
    \label{fig5}
\end{figure}

\begin{figure}
     \centering
     \begin{subfigure}[b]{0.45\textwidth}
         \centering
         \includegraphics[width=\textwidth]{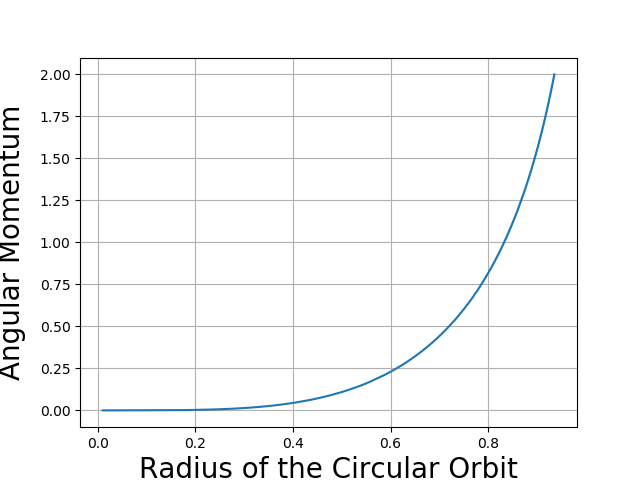}
         \label{fig6a}
     \end{subfigure}
     \hfill
     \begin{subfigure}[b]{0.5\textwidth}
         \centering
         \includegraphics[width=\textwidth]{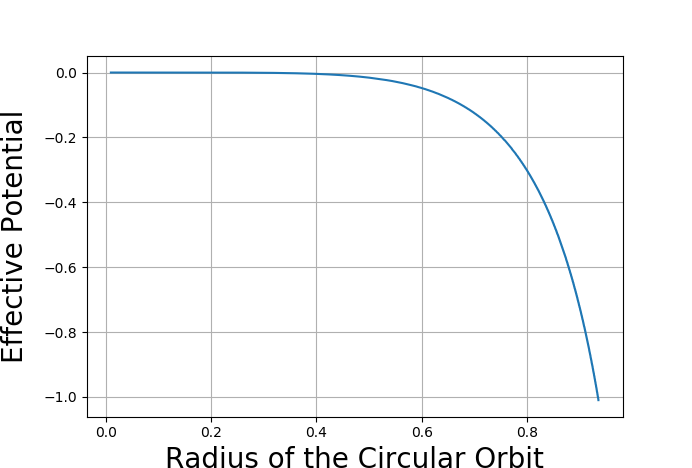}
         \label{fig6b}
     \end{subfigure}
        \caption{The figure on the left shows the precise angular momentum required for a given radius of the circular orbit; the figure on the right shows the correspondent effective potential, which is negative.}
        \label{fig6}
\end{figure}

The first case to consider is the extremal case ($\epsilon, Q=1$), where the potential only depends on $L$, and the relation is illustrated in Figure \ref{fig5}.
As the angular momentum decreases, the concave curve flattens. When there is no angular momentum ($L=0$), the potential vanishes ($V=0$ at all radii). This is known as the stacking property of extremal branes, when the two extremal branes are parallel and stationary, they can not detect the presence of each other. The mass density and charge density are equal and uniform on the brane, thus the gravitational attraction is offset by the electric repulsion. Otherwise, to have a circular orbit at a certain radius requires the angular momentum to be exactly the solution of $L$ from  \ref{6.16}. By substituting $L$ back to \ref{6.15}, there is only a specific value of potential $V_{\mbox{min}}$ for a given circular orbit.  For a given $r$, the radius $L$ and the potential $V$ are solved numerically and their relationship with radius is shown in Figure \ref{fig6}.

In the \ref{fig6}, the radius is bounded from above, implying that outside of a critical radius, there no longer exists a stable circular orbit. The range of radius that can support a circular orbit is $0<r<r_{\mbox{critical}}$. This is due to the required $L$ increasing dramatically as $r$ increases, and the angular momentum is restricted by special relativity. In the case of a Schwarzschild black hole or a Reissner-Nordstrom black hole, there is an inner bound for a stable circular orbit, while in the brane case it becomes an upper bound. This difference is caused by two factors: firstly, the brane does not have a true singularity at $r=0$; secondly, the brane takes $V_{-}$ instead of $V_{+}$ solution to satisfy the boundary condition.

In the extremal case, the asymptote of the potential at radial infinity is zero, $V_{r\xrightarrow{}\infty}=0$. Combined with the fact that $V_{r\xrightarrow{}\infty}$ implies that there always exists a circular orbit given angular momentum. For a conventional Reissner-Nordstrom black hole, as the value of $\epsilon Q$ decreases below one, the concavity of the curve decreases. When $\epsilon Q$ is further reduced to negative (the charge of the probe and the black hole are opposite),  the potential can no longer support a circular orbit. 

\begin{figure}[htp]
    \centering
    \includegraphics[width = 0.60\textwidth]{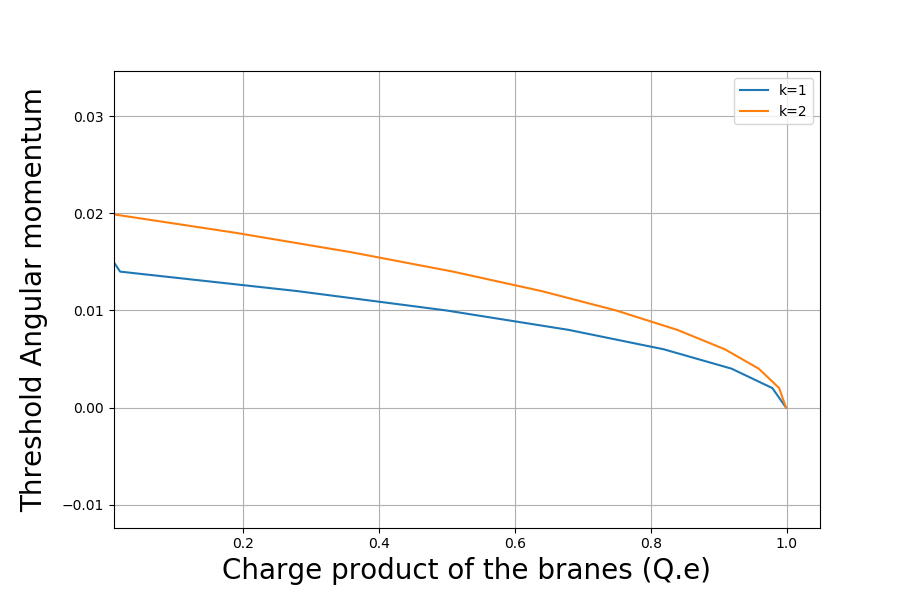}
    \caption{Effective potential under different angular momentum}
    \label{fig7}
\end{figure}
The $0 \leq \epsilon Q < 1$ corresponds to the non-extremal case with the presence of black branes.
The last possible circular orbit is where the asymptote in the non-extremal case is $\epsilon Q - 1$ is equal to the minimum point, i.e. $V_{min}(\epsilon, Q, L)=\epsilon Q - 1$. Because the exact expression of $V_{min}(L)$  can not be easily determined, the numerical method is devised to investigate. The asymptote of the potential $V_{\infty}(\epsilon Q)$ and the minimum point of the potential $V_{\infty}(L=0.01,\epsilon Q)$ are plotted against $\epsilon Q$, and the interception of the two gives the value $\epsilon Q$, below which a stable circular orbit can not exist for a probe with $L=0.01$. Therefore, for a given value of $\epsilon Q$, the code intends to find the threshold angular momentum $L_{\mbox{threshold}}$, beyond which a stable circular orbit always exists. The relationship between the $L_{\mbox{threshold}}$ and $\epsilon Q$ is shown in \ref{fig7}

As shown in \ref{fig7}, for $k=1$ and $k=2$, when $L_{\mbox{threshold}}$ is above 0.014 and 0.020, respectively, it requires $\epsilon Q$ to be negative to have a circular orbit. Because the black brane is assumed to have the same sign of charges $0 \leq \epsilon Q$. This implies if the probe has an angular momentum above 0.014 for $k=1$ or 0.020 for $k=2$, it will always have a circular orbit.

\section{Non-Circular Orbit}
To obtain circular orbits, only the $x^m=t$ and $x^m=\theta$ components are used from the equation of motion in \ref{5.23}. In order to obtain the generalised orbit the $x^m=r$ component is required, and different to the previous two, which simply yields two conserved quantities, all the field content ($g_{mn}$ and $\phi$) are dependent on $r$. Consequently, the equation of motion of $x^m=r$ is a complicated non-linear differential equation shown as followed
\begin{equation}\label{6.17}
\begin{aligned}
      \ddot{r}+\dot{x^n}\dot{x^p}\Gamma^r_{np}&=\partial^r\phi_2-\dot{r}^2\partial_r\phi_2+e^{-\phi_2} \partial^r{H^{-1}}\dot{t}\\
    \ddot{r}-\frac{1}{2}\dot{\theta}\dot{\theta}g^{rr}g_{\theta\theta,r}-\frac{1}{2}\dot{t}\dot{t}g^{rr}g_{tt,r}+\frac{1}{2}\dot{r}\dot{r}g^{rr}g_{rr,r}&=\partial^r\phi_2-\dot{r}^2\partial_r\phi_2+e^{-\phi_2} \partial^r{H^{-1}}\dot{t}.
\end{aligned}
\end{equation}
After making the following substitution
\begin{equation}
    \begin{aligned}
    g_{tt}&=-H^{-6/7}\\
    g_{rr}&=H^{1/7}\\
    g_{\theta\theta}& =H^{-1/7} r^2\\
    g^{rr}&=H^{-1/7}\\
    \dot{t}&=H^{10/7}(\mathcal{E}-\epsilon Q H^{-1})\\
    \dot{\theta}&=H^{3/7}\frac{L}{r^2}\\
    e^{\phi_2}&=H^{-4/7}
\end{aligned}
\end{equation}
\ref{6.17} will become a differential equation of with only one variable $r$. Thereby, one can insert solution of this differentiation $\dot{r}$ back to \ref{6.13} and obtain the effective potential $V$. Due to the complexity of the algebra, the non-circular orbit is not a focus of this project.

\chapter{Conclusion}
In this review, the $D=11$ supergravity and the properties of the p-brane ansatz have been explored. The $D=11$ supergravity model contains one fermionic field, the gravitino, and its super-partner, gravity and the antisymmetric tensor field. Together they form a theory with local supersymmetry where all the interactions are governed by the action in \ref{3.1}. 

The focus is then turned to the bosonic sector of the action in \ref{3.3}. Supergravity is regarded as an effective theory of string theory in the low energy limit, which is the primary motivation to explore supergravity and its solutions. The correspondence between the two is described in Section \ref{sec3.1}: $D=11$ supergravity can be dimensionally reduced to $D=10$, which takes the same form as $D=10$ $\sigma$-model. The dilaton in the $\sigma$-model corresponds to a component in the metric in supergravity, and this scalar contribution is kept in the single-charge action in \ref{3.25}. 

The field equation is derived by varying the single-charge action is derived, shown in \ref{3.31}. Due to the non-linear nature of the field equation, a symmetric ansatz is proposed. The ansatz decomposes spacetime into a Minkowski world volume for the p-branes and an isotropic transverse space. From this ansatz, all fields in the theory only have radial dependence $\left(\phi(r), g_{MN}(r), F_{[4]}(r)\right)$, which solution is shown in \ref{3.60}. The solution bifurcates into an electric 2-brane and a magnetic 5-brane, where the former is dynamical defined by the gauge field and the latter is topological defined by the field strength. 

The p-brane ansatzes are phenomenologically similar to black holes.
The metric of the brane appeared to have a naked singularity, but the curvature at the singularity does not diverge. After a coordinate transformation, the metric becomes Schwartzchild like, such that the true singularity is covered by an event horizon. In fact, the brane can be dimensionally reduced to charged black holes in the transverse space. For a Riessner-Nordstrom black hole, it has two horizons unless being in the extremal case, in which they coincide. The branes only have one horizon, which implies the ansatz automatically saturates the BPS bound, where the charge and mass densities are equal. The charge density is calculated through a Gauss integral over the boundary of the transverse space; the mass density is calculated through the ADM formalism. The two quantities are proportional to the volume-element of the transverse space and are indeed equal. 

The motion of a small probe brane around a parallel massive brane is also similar to orbital motions around a black hole. After dimensional reduces the spatial part of the electric brane's world volume, the action of the brane is the same as the action of a particle around an extremal Riessner-Nordstrom black hole (up to the scalar). The parallel branes can remain stationary with no potential in between because the gravitational attraction offsets the electric attraction
The circular motion of the branic orbit is then investigated through a combination of analytical and numerical methods. The circular orbit with a certain radius corresponds to a precise angular momentum in the extremal case. The non-extremal case can be achieved by black brane, which can exist with another ansatz, in which case the circular orbit would be sensitive to the charge-mass ratio of the brane.

To solve the effective potential of general non-circular orbit, the radial component $x^m=r$ of the field equation needs to be used. The equation is non-linear, and thus requires more advanced computational methods or analytical ingenuity to solve. The general orbit is beyond the scope of this project, nevertheless, it can be an avenue to explore in the future.

\bibliographystyle{ieeetr}
\bibliography{sample.bib}

\end{document}